\pdfoutput=1
\documentclass[aps,prd,10pt,preprintnumbers,showpacs,twocolumn,
superscriptaddress , floatfix,nofootinbib]{revtex4-1}

\usepackage{latexsym} 
\usepackage{amssymb} 
\usepackage{amsmath} 
\usepackage{amsfonts}
\usepackage{bm}
\usepackage{color}
\usepackage{times} 
\usepackage{hyperref}
\usepackage[utf8x]{inputenc}
\usepackage{graphicx} 
\usepackage{siunitx}
\usepackage[inline]{enumitem}
\usepackage{xspace}

\graphicspath{ {../plots/} }

\newcommand{\Eref}[1]{Eq.~\eqref{#1}} 
\newcommand{\Fref}[1]{Fig.~\ref{#1}} 
\newcommand{\Sref}[1]{Sec.~\ref{#1}} 
\newcommand{\Tref}[1]{Table~\ref{#1}}

\begin{document}

\title{Modern tools for computing neutron star properties}

\author{Wolfgang Kastaun}
\author{Frank Ohme}

\affiliation{Max Planck Institute for Gravitational Physics
(Albert Einstein Institute), Callinstr. 38, D-30167 Hannover, Germany}
\affiliation{Leibniz Universität Hannover, D-30167 Hannover, Germany}

\begin{abstract}

Astronomical observations place increasingly tighter and more diverse
constraints on the properties of neutron stars (NS).
Examples include observations of radio or gamma-ray pulsars, 
accreting neutron stars and x-ray bursts, magnetar giant flares, 
and recently, the gravitational waves (GW) from coalescing binary
neutron stars. 
Computing NS properties for a given EOS, such as mass, radius, 
moment of inertia, tidal deformability, and innermost stable circular 
orbits (ISCO), is therefore an important task.
This task is unnecessarily difficult because relevant formulas are 
scattered throughout the literature and publicly available 
software tools are far from being complete and easy to use.
Further, naive implementations are unreliable in numerical  
corner cases, most notably when using equations of state (EOS)
with phase transitions.
To improve the situation, we provide a public library for 
computing NS properties and handling of EOS data. Further, we include a 
collection of EOS based on existing nuclear physics models together
with precomputed sequences of NS models. All methods are accessible 
via a Python interface.
This
article collects all relevant equations and numerical methods
in full detail, including a novel formulation for the tidal 
deformability equations suitable for use with phase transitions.
As a sidenote to the topic of ISCOs, we discuss the stability of 
non-interacting dark matter particle circular orbits \emph{inside} NSs.
Finally, we present some simple applications relevant for parameter 
estimation studies of GW data. For example, we explore the 
validity of universal relations, and discuss the appearance 
of multiple stable branches for parametrized EOS. 
\end{abstract}

\pacs{
04.25.dk,  04.30.Db, 04.40.Dg, 97.60.Jd, }

\maketitle

\section{Introduction}
\label{sec:intro}

Neutron stars (NS) are among the most interesting astrophysical objects,
as their description requires both general relativity and nuclear 
physics. 
The latter comes into play via the equation of state (EOS) of
matter up to densities exceeding nuclear saturation density.
The EOS is assumed to be universal, i.e., not dependent 
on the star's origin.
NSs are distinguished individuals with different 
surface temperature, magnetic field strengths and 
topologies, rotation rates, and propensity for glitches
or radiation outbursts.
However, they can be described by a few scalar properties 
well enough for many astrophysical applications. 
The most basic NS features are captured by the simple case of
slowly rotating, non-magnetized stars.
Such models are completely determined
by the central density and the EOS.
The purpose of this paper and the provided software is to 
enable readers to compute those models with ease.

The most important NS properties are the following.
First and foremost, the gravitational mass, which completely 
determines the metric outside the NS. 
There is also a ``baryonic mass'', which expresses 
the baryon number in units of mass, and which is important 
as a conserved quantity, e.g., in the context of neutron star mergers.
The difference between baryonic and gravitational mass 
defines the binding energy.
The NS size is usually expressed as the proper circumferential 
radius, which also determines the surface area. 
Another important property is the compactness, defined as 
gravitational mass over proper circumferential radius.
It directly determines the surface redshift, and is strongly 
correlated with other properties such as oscillation 
frequencies, moment of inertia, or tidal deformability.
The moment of inertia is obviously relevant for questions
related to spin-down or glitches. 
It also determines the lowest order rotational 
corrections to the metric in comparison to the nonrotating case.
The tidal deformability of a NS is the proportionality factor
between external tidal fields and the induced quadrupole moment.
It is very important for the late inspiral phase of BNS mergers
as it determines the main corrections of the orbital dynamics
compared to the binary black hole case, and thus the 
observable gravitational wave (GW) signal.

The equations governing mass and radius of nonrotating 
neutron stars (NS) are known since the early 
work by \cite{Tolman:1939, Oppenheimer:1939}. 
This also established that the NS mass is bounded,
and that the maximum mass depends on the EOS.
Not all solutions for static NS are stable against
radial perturbations. The single parameter sequence 
of NS consists of
one or more stable and unstable branches, which 
depends on the EOS.
Criteria for the stability have been collected by 
\cite{Bardeen:1966:505B}.
The equations for the moment of inertia were derived by 
\cite{Hartle:1967} for slowly rotating NS. More recently,
equations that govern the tidal deformability have been derived in 
\cite{Hinderer:2008:533487,Hinderer:2008:533487:erratum,Flanagan:2008:021502,Hinderer:2010:123016}.
The computation of the above properties requires the 
solution of an ordinary differential equation system (ODE) 
with singular boundary conditions.

One complication 
is given by the possibility of phase transitions in the EOS,
which can lead to discontinuities in the energy density as function
of pressure. We note that phase transitions do not necessarily lead 
to discontinuities. Those which do are the problematic ones 
in the context of this work, and we will use the 
terms synonymously in the following.
Although a true discontinuity can be treated analytically for the tidal 
deformability ODE
\cite{Postnikov:2010:024016, Takatsy:2020:028501},
such treatment is infeasible for the more typical case of EOS that merely 
exhibit very sharp features. That case is also problematic 
for direct numerical solution.

Measurements of NS properties can help
to constrain the equation of state (EOS) of neutron star matter.
There are different avenues towards this goal. 
Any measurement of a NS mass provides a lower bound for the 
maximum NS mass.
Another avenue is the measurement 
of mass-radius relations by electromagnetic observations (see, e.g., 
\cite{Oezel:2010:101301,Oezel:2016:023322,Oetzel:2006:04858}). 
The moment of inertia can also serve to constrain the EOS and might
be measured, e.g., via observation of double pulsar systems
\cite{Raithel:2016:032801}.
Another possibility is the measurement of mass and tidal deformability
using observations of gravitational waves from BNS coalescence, 
such as the famous event GW170817 
\cite{LVC:BNSDetection,LVC:BNSSourceProp:2019,LVC:GWTC1:2019, LVC:MMA:2017,AdvLIGO:2015,Virgo2015}.
Further, the stability of a BNS merger remnant is related
to the maximum NS mass. If electromagnetic counterparts in
a BNS multimessenger observation carry information on the fate of the 
remnant, it can be used to constrain the EOS. This was already
done (under additional assumptions) using the short gamma ray burst 
associated with GW170817 \cite{LVC:GWGRB:2017}.

Despite the astrophysical importance of computing NS properties,
the available software infrastructure is very limited. Although
there exists a plethora of solvers for the basic NS structure,
these codes are severely lacking with regard to some of 
the following aspects: 
\begin{enumerate*}[label={\alph*)}]
\item public availability
\item ease of installation
\item documentation
\item dependence on free open source software only
\item reliability, also in corner cases
\item error estimates
\item code quality
\item usability from within other codes
\item completeness of NS properties and
\item EOS handling.
\end{enumerate*}

One purely technical hurdle is the lack of a standardized 
exchange format for generic EOS.
One notable file format for tabulated nuclear physics EOS 
is developed by the CompOSE project \cite{ComposeEOS}.
It is well-documented and complete in the sense that all 
required metadata is contained. However, this standard 
does not allow popular analytic EOS models,
leaves the interpolation method unspecified, and lacks an 
easy to use interface for reading and evaluating an EOS 
from within other code. 

The aim of our work is to address all of the above issues.
Recently, we provided tools for computing NS properties as
part of the library \texttt{RePrimAnd}, which was developed 
to support general relativistic magnetohydrodynamics 
simulations \cite{Kastaun21:023018}.
It also provides an elaborate framework for handling of EOS.
The library is publicly available and documented \cite{RePrimAnd:v1.7}.
It can be used from within C++ or Python.

This article collects everything required for computing NS 
properties in the \texttt{RePrimAnd} library, such as basic 
notation and definitions,
all equations needed for computing NS structure, 
and a discussion how to avoid numerical pitfalls.
We convert formulas scattered across the literature 
into a coherent notation and point out details 
that are usually not discussed but important for actual
computations.
Further, we reformulate
some equations to facilitate numerical solution. 
Most notably, we provide a formulation of the differential 
equations for tidal deformability that is robust
when the EOS exhibits phase transitions, and we correct
a faulty approximation for the limit of low compactness.

We also perform extensive tests of the accuracy, which leads 
to a model for the error bounds. This model is incorporated 
in the library, allowing to directly specify the desired 
accuracy. A related practical problem regards the impact
of approximated EOS representations that use interpolation 
and low-density extrapolation. We will provide some guidance 
regarding requirements for tabulated EOS.

Finally, we present some simple results obtained with the new
library, which might be useful for gravitational wave (GW) 
and multi-messenger astronomy.
Firstly, we collect NS properties for a number of nuclear 
physics EOS available in the literature, such as tidal 
deformability and moment of inertia.
Next, we study the reliability of empirical relations
between NS compactness and tidal deformability that 
allegedly depend only weakly on the EOS, by
performing numerical searches for EOS causing larger 
deviations.
The finding are relevant for studies that combine constraints
of mass-radius and mass-tidal deformability relations,
e.g., using NICER and GW data.
We also point out how splitting of stable branches 
complicates parameter estimation with parametrized 
EOS for BNS GW detections.
Last but not least, we provide a small collection of 
ready-to-use EOS files representing existing nuclear physics 
EOS models.

\section{Formulation}
\label{sec:formulation}
In the following we collect all equations needed for computing
the properties of nonrotating NS. Most are well known but scattered
throughout the literature and use different notation and conventions.
We recast some expressions into the variants that are employed in the 
\texttt{RePrimAnd} library, and which are advantageous for numerical
solution. The main novelty, described in \Sref{sec:tidalode}, is our 
new formulation of the ODE that governs tidal deformability, which
is applicable also to EOS with phase transitions.  
\subsection{Conventions}
\label{sec:notation}

We use geometric units throughout this work unless noted otherwise, 
that is, units in with $G=c=1$.
This leaves open one degree of freedom, which can be fixed by choosing 
a mass unit. Denoting time, length, and mass units by $u_T, u_L, u_M$, 
respectively, the unit system is given by
\begin{align}
u_L &= u_M \frac{G}{c^2}, & 
u_T &= \frac{u_L}{c}
\end{align}

We stress that when computing NS properties using equations assuming 
geometric units,
the gravitational constant $G$ is implicitly given by the geometric 
unit system. Comparing results obtained in different geometric unit 
systems is not just a matter of converting back to SI units, unless
both use the same value of $G$. In contrast, the choice of the mass 
unit is irrelevant because, unlike the constant $G$, it is not a 
physical constant that would appear in the full equations stated in 
arbitrary units. 
Often, geometric unit systems are specified by the condition 
$G=c=M_\odot=1$. Without providing the precise values 
assumed for $G$ and $M_\odot$, it is impossible to compare results 
to a precision better than around $0.01\,\%$ since both constants 
are not known very accurately (although their product is).
For the NS solutions in this work, we use values of exactly
$G=\SI{6.67430E-11}{\meter\cubed\per\kilogram\per\second\squared}$ and
$M_\odot=\SI{1.98841E30}{\kilogram}$ (from \cite{PDG:Constants:Zyla:2020zbs}).

We assume that NS matter in GR can be described by the stress-energy tensor
of a perfect fluid, given by
\begin{align}
T_{\mu\nu} &= (E+P) u_\mu u_\nu + P g_{\mu\nu}
\end{align}
Above, $u$ is the 4-velocity of the fluid, 
$E$ is the total energy density in the restframe of 
the matter, and $P$ is the pressure.
This means that we make the approximation of isotropic pressure, 
and exclude any shear-stresses.
For a discussion of NS structure with anisotropic pressure, we 
refer to \cite{Pretel:2023:18770P:arXiv}.
We note that realistic NS can 
have minor shear stresses within the crust. This is irrelevant
when computing spherical equilibrium models under the assumption
of zero shear deformation, and it is commonly neglected when computing 
the tidal deformability.

We denote the baryon number density in the fluid restframe by
$n_B$. Introducing a mass constant $m_B > 0$, one can define 
a ``baryonic mass density'' (or mass density for short) 
$\rho \equiv m_B n_B$.
The constant $m_B$ is arbitrary and usually chosen around the 
neutron mass, 
while the exact value varies between different sources.
When comparing baryonic mass density between different sources,
this should be taken into account.
In the \texttt{RePrimAnd} framework, the convention is 
$m_b=\SI{1.66E-24}{\gram}$.

Further, we will use the 
specific internal energy $\epsilon$ and
relativistic enthalpy $h$ 
defined by
\begin{align}
\epsilon &\equiv \frac{E - \rho}{\rho} \\
h &\equiv  \frac{E+P}{\rho} = 1 + \epsilon + \frac{P}{\rho}
\end{align}
Note that the  definitions of $\rho, \epsilon, h$ depend on 
the choice of $m_B$, such that $\rho$ and $h$ differ by a global 
factor between different choices. The specific internal energy 
$\epsilon$ also differs by an offset, not just a factor.
In particular, $E \rightarrow 0$ for $\rho \rightarrow 0$ does not 
imply that $\epsilon$ approaches zero, nor that it is positive.
The zero-density limit depends on the choice of $m_B$. 
The conditions $\epsilon>-1, h>0$ hold independently of this choice, 
assuming only that $E >0, P \ge 0$.

\subsection{Equation of State}
\label{sec:eos}
In general, NS matter has three degrees of freedom,
usually parametrized in terms of baryon number density $n_B$, 
temperature $T$, and electron fraction $Y_e$ (one usually 
assumes macroscopic charge neutrality since astrophysical 
objects are assumed to carry negligible net charge, and NS 
matter is assumed to be highly conductive). 
The variables $n_B$ and $Y_e$ are equivalent to the 
thermodynamic state variables $V$ (volume) and 
particle numbers $N_i$ for protons and neutrons.

The behavior of 
matter is completely described by a single thermodynamic 
potential, meaning a scalar function of a particular set of  
state variables. There are different but completely equivalent 
thermodynamic potentials, each based on a different set of 
state variables.
When using state variables $(V,T,N_i)$, the corresponding 
potential is the Helmholtz free energy $F(V,T,N_i)$. 

Given a thermodynamic potential and its canonical state variables, 
one can derive all other quantities. 
Such derived relations are collectively referred to as 
equation of state (EOS).
For the Helmholtz free energy, 
pressure and entropy are given by the partial derivatives 
$P=\partial F / \partial V$ and 
$S=\partial F / \partial T$, respectively, 
and the internal energy is given by $U=F+TS$.
We can parametrize the EOS as functions
$P(\rho, T, Y_e)$, $E(\rho, T, Y_e)$, and 
$S(\rho, T, Y_e)$.

For most applications, only the EOS is needed in some form,
but not the underlying thermodynamic potential. In fact,
nuclear physics models are often distributed as tables 
sampling the EOS functions. 
We note that this makes it more difficult to interpolate such 
tables in a consistent manner.
There are also some toy models where the EOS is directly prescribed
as analytic expressions.

It should be noted that one cannot chose the various EOS 
functions independently. The existence of a thermodynamic potential 
implies thermodynamic consistency constraints. For example, 
$\partial P / \partial T = \partial S / \partial V 
= \partial^2 F / \partial T \partial V$.
If those constraints are violated, an EOS is physically invalid.
Results based on such inconsistent EOS are not just wrong, but 
ambiguous, since there are different ways to express the same 
quantity which cease to yield identical results.

In this work, we are concerned only with scenarios where
pressure and energy density can be expressed as functions 
$P(\rho)$ and $E(\rho)$ of the mass density alone.
This is called a barotropic EOS. 
Physically, employing a barotropic EOS means that
two of the matter degrees of freedom are restricted somehow.
The most relevant example for our aims is to model cold
neutron stars, where thermal effects can be neglected and 
one can set $T=0$ for all practical purposes.
Further, we want to model equilibrium models. For those, the 
electron fraction becomes a function of density because weak 
processes drive the matter towards $\beta$-equilibrium. 

For perturbed NS,
$\beta$-equilibrium is maintained if the perturbation timescale
is much longer than the intrinsic timescales of weak processes.
This is the case for tidal deformations in the limit of large 
separation. 
We note that for the timescales accessible to ground-based 
gravitational wave detectors, approximating deformations as
static might be insufficient in any case. 
For further discussion of dynamical tidal effects see, e.g., 
\cite{Dietrich:2017feu}. 
Here, we only consider static deformations.

In this work, we assume that pressure cannot be negative, 
and we exclude exotic types of matter, assuming that $E \ge 0$ and $\epsilon>-1$.
We further exclude any contributions to energy or pressure
unrelated to baryonic matter, such as dark matter clouds
or radiation fields outside a NS. Thus, we assume
$P(0) = E(0) = 0$.

An important subset of barotropic EOS is given by the
isentropic barotropic EOS, which are defined by a constant
specific entropy. A NS model following such an EOS will continue 
to do so when perturbed adiabatically, i.e., 
such that $T \mathrm{d}S = 0$ and hence 
$P \mathrm{d}V + \mathrm{d}U = 0$.
For barotropic EOS, this condition implies that
\begin{align}\label{eq:eos_adiabatic}
\frac{\mathrm{d} E}{\mathrm{d}\rho} &= h, &
\frac{\mathrm{d} \epsilon}{\mathrm{d}\rho} &= \frac{P}{\rho^2}, &
\frac{\mathrm{d} h}{\mathrm{d}P} &= \frac{1}{\rho}
\end{align}

Our main use case are cold NSs, which indeed follow an isentropic 
barotropic EOS. 
One can construct other examples of NS with isentropic barotropic 
EOS by making the artificial assumption of a radial temperature 
profile such that the specific entropy remains constant.
A counter-example that cannot be modeled that way is a 
hot NS with constant temperature.

How matter reacts to small perturbations 
can be expressed by the adiabatic speed of sound, which for an 
isentropic barotropic EOS is given by the derivative
\begin{align}\label{eq:eos_csnd_adiab}
c_s^2 &= \frac{\mathrm{d} P}{\mathrm{d} E} 
       = \frac{1}{h}\frac{\mathrm{d} P}{\mathrm{d} \rho}
\end{align}
On physical grounds, we rigorously demand that any EOS 
satisfies
\begin{align}\label{eq:eos_cs2_limits}
0 &\le c_s^2 < 1
\end{align}

The condition $c_s^2 \ge 0$ is required for stability of matter,
otherwise perturbations would be exponentially growing instead
of propagating as sound waves. This condition sometimes gets 
violated when stitching 
together EOS computed for different density regimes, or when 
interpolating coarsely sampled EOS using ill-suited interpolation 
methods that produce overshoots.

The causality condition $c_s < 1$ is required for any physically 
valid model. Already on the mathematical level, the equations 
that govern relativistic hydrodynamics break down if there are 
superluminal characteristic speeds. 
Note that the equations describing the static solutions discussed 
here do admit mathematically valid solutions also for the case 
with superluminal soundspeeds. 
However, we do not consider such solutions since they are 
not valid in the wider context of the general relativistic 
hydrodynamics evolution equations.

We point out that nuclear physics EOS models are based on
approximations and often violate causality above some 
density. We still use portions
of such EOS, restricting the validity range to lower densities 
such that $c_s<1$ is satisfied.

For isentropic barotropic EOS, the condition $c_s^2 \ge 0$
implies that $P(\rho)$ increases monotonically but not 
necessarily strictly monotonic. We can therefore obtain the 
inverse function $\rho(P)$, which is strictly monotonic 
but may have discontinuities.

Although valid EOS with non-monotonic $P(\rho)$ could be 
constructed in the non-isentropic case, we exclude such
EOS for the purpose of this work.
The reason is that otherwise there would be an infinitude of 
solutions for spherical NS equilibrium models at given central 
density, caused by the additional freedom of choosing a 
branch from of a multi-valued $\rho(P)$ in any pressure 
interval within the non-monotonic range. We are unaware of an 
astrophysical use case justifying such complications.

Similarly to ordinary matter such as water, nuclear matter may exist 
as a mixture of different phases at the same pressure but different 
density. How nuclear matter behaves while transitioning through such 
a regime depends on the exact nature of those phases. One possibility
is that the pressure as a function of density stays constant within 
the density range of the transition. In this case, the speed of sound 
is zero over the same range (see \Eref{eq:eos_csnd_adiab}). 
The density as function of pressure thus has a discontinuity at such 
a phase transition.

Another possibility is that the pressure does increase across 
the phase transition. This may occur for complex matter with more than 
one conserved charge \cite{Glendenning:1992:1274G}.
The jump in the function $\rho(P)$ induced
by phase transitions can thus vary both regarding its steepness 
and its size (for a discussion of the influence of the steepness 
on NS properties, see \cite{Han:2019:083014}).

We remark that non-isentropic barotropic EOS might have a range where
$P(\rho) = \text{const}$ even if there is no physical phase
transition. For example, one could prescribe specifically designed
functions $T(\rho)$ and/or $Y_e(\rho)$. 
Further, we remark that an EOS with a phase transition might only be
a valid description of matter on timescales longer than
NS oscillation periods. 
This affects the criteria for the stability of NS,
as discussed in \cite{Rau:2023:103035}. 

In the remainder of this work, we will not distinguish the 
different physical scenarios above, since our only concern is 
the impact of discontinuities
or steep gradients on the numerical solutions. We 
will therefore use the term phase transition synonymously for 
any sharp features in the EOS where $c_s \ll 1$ over a density 
range.

One important quantity for equilibrium models is the
pseudo-enthalpy defined as
\begin{align}\label{eq:eos_pseudoh}
H(P) &= \exp\left(\int_0^P \frac{\mathrm{d}P'}{P' + E(P')} \right)
\end{align}
We require that the above integral is finite, such that $H(0)=1$.
This mild restriction on the EOS is not a practical concern.
We note that the enthalpy $h$ depends on the choice of the formal 
baryon mass constant $m_B$, while the pseudo-enthalpy $H$ does not.

We can use $H$ to parametrize the EOS as $P(H)$.
By construction, $H(P)$ is a smooth and strictly monotonic.
Thus, $P(H)$ is also smooth and strictly monotonic. 
This is still true across a phase transition.
Since $P(\rho)$ has a plateau across a phase transition,
the same holds for $H(\rho)$. 
Correspondingly, mass density $\rho(H)$ and energy density 
$E(H)$ have a discontinuity at a phase transition.

The pseudo-enthalpy $H$ obeys the following identity, which is
useful in the context of hydrostatic equilibrium.
\begin{align}\label{eq:eos_dlnhdp}
\frac{\mathrm{d} }{\mathrm{d} P} \ln H(P)
&= \frac{1}{E(P) + P}
\end{align}

For isentropic barotropic EOS, the pseudo enthalpy $H$ agrees with the 
regular enthalpy $h$ up to a constant factor, that is, 
\begin{align}\label{eq:eos_isentropic_h_at_rho}
H(\rho) &= h(\rho) / h(0).
\end{align} 
This can be shown by combining 
\Eref{eq:eos_adiabatic} and \Eref{eq:eos_dlnhdp} to obtain 
$\mathrm{d} \ln(H) / \mathrm{d} \ln(h) = 1$.
For isentropic EOS, we can also write \Eref{eq:eos_csnd_adiab} as
\begin{align}\label{eq:eos_csnd_adiab_alt}
c_s^2 &= \frac{\mathrm{d} \ln\left(H\right)}{\mathrm{d} \ln \left(\rho \right)} 
\end{align}

\subsubsection{Polytropic EOS}

The polytropic EOS (or polytrope for short) is a barotropic 
isentropic EOS that is frequently
used in the context of neutron stars as a toy model for reference. 
We remark that polytropes already appear in classical thermodynamics 
as curves of constant specific entropy for the classical ideal gas EOS. 
That is not how they are used in 
the context of nuclear matter, however. For the classical ideal gas,
the pressure at zero temperature is zero, whereas cold nuclear matter 
has the degeneracy pressure arising from Pauli's exclusion 
principle and other contributions.  
Polytropic EOS can still be used as a simple analytic prescription
to approximate zero-temperature nuclear matter. When used like 
this, they are not derived from any thermodynamic potential. 
Also, being a toy model, polytropic EOS neither depend on 
nor provide the electron fraction.

The polytropic EOS is given by
\begin{align}\label{eq:eos_poly_p_at_rho}
P(\rho) &= K \rho^\Gamma, & \Gamma &\equiv 1 + \frac{1}{n}
\end{align}
The constant $\Gamma$ is a parameter called polytropic exponent and
is alternatively specified by the polytropic index $n$.
The constant $K$ is called polytropic constant and it has 
awkward units with non-integral exponents depending on $\Gamma$. 
We therefore use an alternative constant $\rho_p$ with units of 
a density, writing
\begin{align}\label{eq:eos_poly_p_at_rho_alt}
P(\rho) &= \rho_p \left(\frac{\rho}{\rho_p}\right)^\Gamma, 
& \rho_p &\equiv K^{-n}
\end{align}
The specific energy follows from the adiabatic assumption 
\Eref{eq:eos_adiabatic}
\begin{align}\label{eq:eos_poly_eps_at_rho}
  \epsilon(\rho) 
  &= \epsilon_0 + n \left( \frac{\rho}{\rho_p} \right)^\frac{1}{n} 
\end{align}
The constant $\epsilon_0$ is another free parameter, although it is 
typically set to zero.
Enthalpy and pseudo-enthalpy follow as
\begin{align}\label{eq:eos_poly_h_at_rho}
  H(\rho) &= \frac{h}{h_0} = 1
             + \frac{ n + 1 }{h_0} \left( \frac{\rho}{\rho_p} \right)^\frac{1}{n}
  ,& h_0 &= 1+ \epsilon_0 
\end{align}
We can parametrize the EOS in terms of $H$ as follows
\begin{align}
  P(H) &= \rho_p \left(\left(H-1 \right) \frac{h_0}{1+n} \right)^{1+n} 
  \label{eq:eos_poly_p_of_h}\\
  \epsilon(H) &= \epsilon_0 + \frac{h_0}{\Gamma} \left( H-1 \right) 
  \label{eq:eos_poly_eps_of_h}\\
  \rho(H) &= \rho_p \left(\left(H-1 \right) \frac{h_0}{1+n} \right)^n
  \label{eq:eos_poly_rho_of_h}
\end{align}
Finally, the soundspeed follows from \Eref{eq:eos_csnd_adiab_alt} as 
\begin{align}\label{eq:eos_poly_csnd_of_h}
c_s^2(H) &= \frac{H-1}{n H}
\end{align}
This constrains the range where the EOS is physically valid.
The condition $c_s^2 \ge 0 $ requires that 
$n>0$ (equivalent to $\Gamma>1$). 
Further, we find that $c_s < 1$ for any density if 
$n \ge 1$ (equivalent to $\Gamma \le 2$). For the case $\Gamma > 2$, 
causality is violated above a critical density given by 
\begin{align}\label{eq:eos_poly_h_valid}
H_c &= \frac{1}{1-n}
\end{align}

For comparison between different sources, we should discuss what happens when 
changing between a convention using $m_b$ to using another value $m_b'$.
First, we note that the constant $\rho_p$ does \emph{not} transform like the baryonic 
mass density $\rho$. Instead, the conditions $P'=P, \rho/m_b = \rho' / m_b'$
lead to $\rho_p'/\rho_p = (m_b'/m_b)^{1+n}$. Second, the offset $\epsilon_0$ 
changes as  $(1+\epsilon_0')/(1+\epsilon_0) = m_b'/m_b$ (this also implies that one 
can find a value $m_b$ such that $\epsilon_0=0$).

For many applications, one can ignore the above issue.
The value $m_b$ only enters when computing baryon numbers or number densities.
Using $\rho'=\rho, \rho_p'=\rho_p, \epsilon_0'=\epsilon_0$ instead of the formally 
correct transformation will yield exactly the same results for most NS properties,
including the baryonic mass. Only the definition of baryonic mass changes, such that 
the total baryon number of the NS will differ.

\subsubsection{Joining EOS Segments}

Often it is useful to assemble a barotropic isentropic EOS 
from several parts, using different prescriptions in different 
density ranges. The matching condition is that $P(\rho)$ and 
$E(\rho)$ are 
continuous across the segment boundaries (compare \Eref{eq:eos_adiabatic},\Eref{eq:eos_csnd_adiab}, and \Eref{eq:eos_cs2_limits}). 

One application is to extend an EOS based on sample points with strictly 
positive density down to zero density in a well-defined way. This can 
also be regarded as a way of interpolating between zero density and the
lowest non-zero sample point.
Assuming we have an arbitrary EOS that is defined above some density $\rho_m$,
with pressure $P_m = P(\rho_m)$ and specific internal energy $\epsilon_m = \epsilon(\rho_m)$, we obtain a matching polytropic EOS 
from \Eref{eq:eos_poly_p_at_rho_alt} and \Eref{eq:eos_poly_eps_at_rho} as 
\begin{align}\label{eq:eos_match_poly}
\rho_p &= \rho_m \left(\frac{\rho_m}{P_m}\right)^n, &
\epsilon_0 &= \epsilon_m - n \frac{P_m}{\rho_m}
\end{align}
where the polytropic index $n$ is a free parameter. 
We point out that a valid choice has to obey the constraint
\begin{align}
n &< \frac{(1 + \epsilon_m) \rho_m}{P_m}
\end{align}
This condition ensures that $h > 0$ and 
$c_s^2 > 0$ (compare 
\Eref{eq:eos_csnd_adiab}). In practice, it is not very restrictive 
since for realistic EOS and low matching densities, $P \ll E$. 
It might be a problem however when matching at high densities
or when matching EOS with unusual low-density behavior.

We also note that fixing 
$n$ such that $\epsilon_0=0$ would serve no meaningful purpose. 
The energy per baryon in the 
zero-density limit depends on the assumed nuclear composition, and 
unless the former agrees with the arbitrary constant $m_b$, we find
that $\epsilon(0) \neq 0$.

A second application of EOS matching is to approximate a given EOS
by joining several polytropic segments appropriately. 
Approximations to many nuclear physics EOS models 
by means of piecewise polytropic EOS are provided in \cite{Read:2009:124032}. 

A piecewise polytropic EOS is fully specified
by providing $\rho_p$ and $\epsilon_0$ for the lowest segment, 
the polytropic exponents $\Gamma_i$ for each segment, and the 
densities $\rho_i$ of the segment boundaries. The parameters
$\epsilon_{0,i}$ and $\rho_{p,i}$ of the remaining segments then follow 
from applying \Eref{eq:eos_match_poly} to each segment boundary.
In this work, we also set $\epsilon_0 = 0$ for the lowest segment,
as in \cite{Read:2009:124032}. As discussed for the polytropic EOS, 
this has no consequences except for the baryon number.

To compute $H(\rho)$, one cannot use \Eref{eq:eos_poly_h_at_rho} valid
for polytropic EOS. The reason is that $H$ is an integral quantity,
given by \Eref{eq:eos_pseudoh}, and one has to split the integral into 
the individual segments. However, it is much easier to use the fact 
that, since the piecewise polytropic EOS is isentropic by construction, 
it must fulfill \Eref{eq:eos_isentropic_h_at_rho}. We can therefore use 
the regular enthalpy $h$, computed from \Eref{eq:eos_poly_p_at_rho} 
and \Eref{eq:eos_poly_eps_at_rho}.
In the range of segment $i$, we obtain the EOS in terms of $H$ as
\begin{align}
  h(H) &= H \\
  P(H) &= \rho_{p,i} \left( \frac{H-1 - \epsilon_{0,i}}{1+n_i} \right)^{1+n_i} 
  \label{eq:eos_pwpoly_p_of_h}\\
  \epsilon(H) &= \epsilon_{0,i} + \frac{H-1 - \epsilon_{0,i}}{\Gamma_i}  
  \label{eq:eos_pwpoly_eps_of_h}\\
  \rho(H) &= \rho_{p,i} \left(\frac{H-1 - \epsilon_{0,i}}{1+n_i} \right)^{n_i}
  \label{eq:eos_pwpoly_rho_of_h}\\
  c_s^2(H) &= \frac{H-1- \epsilon_{0,i}}{n_i H}
\end{align}
which is valid for our choice of $\epsilon_{0,0} = 0$.
The same expressions (up to notation) for the piecewise polytropic EOS
can also be found in \cite{Read:2009:124032}.

We note that the valid range of piecewise polytropic EOS can be limited
by causality. For a segment $i$, we find that $c_s>1$ if and only if
\begin{align}
H &\ge \frac{1+\epsilon_{0,i}}{1 - n_i} 
\qquad\text{and}\qquad 
0 < n_i < 1
\end{align}
The lowest segment within which the above condition is satisfied 
then limits the validity range of the entire EOS.

\subsection{TOV Equations}
\label{sec:tovode}
In the following, we collect the equations describing the basic 
structure of spherically symmetric NS, and cast them into the form
used in the \texttt{RePrimAnd} library. 
For a more didactic introduction, 
we refer to textbooks \cite{Shapiro:1983:book,Glendenning:2007:book}.

Any spherically symmetric spacetime metric is static 
and can be written as
\begin{align}\label{eq:tov_line_element}
\mathrm{d} s^2 
&=  - e^{2\nu(r)} \mathrm{d}t^2 + e^{2\lambda(r)} \mathrm{d}r^2 
    + r^2 \mathrm{d}\Omega
\end{align}
For this choice of coordinates, the radial coordinate $r$ is the 
proper circumferential radius.
In this form, the time coordinate is only fixed up to a constant 
factor. We follow the standard choice where coordinate time agrees 
with proper time for an Eulerian observer far away from the star, 
that is, $\nu \rightarrow 0$ for $r \rightarrow \infty$.

The metric potentials $\lambda$ and $\nu$ follow a set of ordinary
differential equations known as TOV \cite{Tolman:1939,Oppenheimer:1939}
equations, which can be written as
\begin{align}\label{eq:tov_dladr}
\frac{\mathrm{d} }{\mathrm{d} r} \lambda(r)
&= r e^{2\lambda(r)} \left( 4 \pi E(r) - \frac{m(r)}{r^3} \right),\\
\label{eq:tov_dnudr}
\frac{\mathrm{d} }{\mathrm{d} r} \nu(r)
&= r e^{2\lambda(r)} \left( 4 \pi P(r) + \frac{m(r)}{r^3} \right),
\end{align}
where
\begin{align}\label{eq:tov_m_of_lambda}
m(r) &\equiv \frac{r}{2} \left( 1 - e^{-2\lambda} \right)
\end{align}
For a didactic derivation we refer to \cite{Glendenning:2007:book}.
\Eref{eq:tov_dladr} can also be written as 
\begin{align}\label{eq:tov_dmdr}
\frac{\mathrm{d} }{\mathrm{d} r} m(r)
&= 4 \pi r^2 E(r) 
\end{align}

The above expressions require the matter state at each radius.
Given the pressure at the center, the pressure elsewhere is related
by the hydrostatic equilibrium condition,
which in turn follows from $\nabla_\mu T^{\mu\nu} = 0$.
\begin{align}
\frac{\mathrm{d} }{\mathrm{d} r} P(r)
&= - \left(E(r) + P(r)\right) \frac{\mathrm{d} }{\mathrm{d} r} \nu(r)
\end{align}

We assume that $E$ can be expressed as function of $P$ alone.
That allows us to integrate the above differential equation.
Using \Eref{eq:eos_dlnhdp}, hydrostatic equilibrium in terms 
of $H$ becomes
\begin{align}\label{eq:tov_dhdr}
\frac{\mathrm{d} }{\mathrm{d} r} \ln H (r)
&= - \frac{\mathrm{d} }{\mathrm{d} r} \nu(r)
\end{align}
Therefore, $H$ can be directly expressed in terms of the metric 
potential $\nu$ as
\begin{align}\label{eq:tov_h_of_nu}
H(r)
&= H(0) e^{\nu(0) - \nu(r)} \equiv H(0) e^{-\mu(r)}
\end{align}

Note that during the numerical ODE integration, we do not need 
the central value $\nu(0)$, only the difference
\begin{align}
\mu(r)
&= \nu(r) - \nu(0)
\end{align}

For our implementation, we use $\mu$ as independent variable
instead of the radius $r$. 
We can obtain
another ODE for the radius by inverting \Eref{eq:tov_dnudr}.
The result is similar to 
\cite{Lindblom:1992:569L}, where $\ln(H)$ is used as independent variable.
However, the ODE is still irregular at the center, which can
be avoided by a simple variable substitution $x\equiv r^2$.
\Eref{eq:tov_dnudr} becomes
\begin{align} \label{eq:tov_dnudx}
\frac{\mathrm{d} \nu}{\mathrm{d} x}
&=  e^{2\lambda} \left( 2 \pi P + \frac{m}{2 r^3} \right) > 0,
\end{align}
which can be inverted, resulting in
\begin{align} \label{eq:tov_dxdmu}
\frac{\mathrm{d}x}{\mathrm{d} \mu} 
&= \frac{\mathrm{d} x}{\mathrm{d} \nu}
=  \frac{2 e^{-2\lambda}}{  4 \pi P + \frac{m}{r^3}}
\end{align}
A similar expression (based on $\ln(H)$) can be found in 
\cite{Postnikov:2010:024016}.
As will be discussed later, the term $m r^{-3}$ has a well-defined
finite limit for $r\rightarrow 0$.
Therefore, the above ODE is completely regular.
The ODE for the metric potential $\lambda$ follows from
\Eref{eq:tov_dladr} and \Eref{eq:tov_dnudr} as
\begin{align} \label{eq:tov_dladmu}
\frac{\mathrm{d} }{\mathrm{d} \mu} \lambda(\mu)
&=  \frac{4 \pi E - \frac{m}{r^3} }{  4 \pi P + \frac{m}{r^3}},
\end{align}
and it is completely regular as well.
The system is closed by the functions $P(H)$ and $E(H)$ 
which are defined by the EOS.

Finally, we need to discuss the boundary conditions.
At the origin, we have $\mu(0)=0$ by definition.
For our choice of radial coordinates, regularity of the metric 
at the center implies $\lambda(0)=0$ 
and, from \Eref{eq:tov_m_of_lambda}, $m(0)=0$.
In order to find the behavior of solutions near the origin, we first
write
\begin{align}\label{eq:split_mbyr3}
\frac{m}{r^3} &= \frac{\lambda}{x} \sigma(-2 \lambda)
\end{align}
where 
\begin{align}
\sigma(l) &= \frac{e^l - 1}{l} \\
&= 1 + \frac{1}{2} l + \frac{1}{6} l^2 + \frac{1}{24} l^3 + \frac{1}{120} l^4
  + \mathcal{O}(l^5) \label{eq:sigma_series}
\end{align}
is a smooth function. 
To get an expression for $\lambda/x$, we
combine \Eref{eq:tov_dxdmu} and \Eref{eq:tov_dladmu} into
\begin{align}\label{eq:dla_dx}
\frac{\mathrm{d}\lambda}{\mathrm{d}x} &= 
\left(4\pi E - \frac{\lambda}{x} \sigma(-2 \lambda) \right) 
\frac{e^{2\lambda}}{2}
\end{align}
Next we employ a Taylor-expansion in terms of $x$, writing
\begin{align}
\lambda &= \sum_i \lambda_k x^k, &
E &= \sum_i E_k x^k
\end{align}
Note that there can be no term linear in $r$ as this would
imply that the final three-dimensional solution
becomes non-smooth at the center of the star.
Inserting the above expansion into \Eref{eq:dla_dx} 
and \Eref{eq:sigma_series} yields
the coefficients. To first order, we find
\begin{align}
\frac{\lambda}{x} &= \kappa_1(E_0, E(\mu), x) + \mathcal{O}(x^2) 
\label{eq:tov_lt_by_r_lim}\\
\kappa_1 &= 
\frac{4\pi}{3} E_0 \left(1 + \frac{3}{5}\left(\frac{E}{E_0} - 1\right) 
                         + \frac{4\pi}{3} E_0 x \right) \label{eq:def_kappa1}
\end{align}
For later use, we also replaced the coefficient $E_1$ by first order 
finite differences in terms of $E,E_0$.
Together with \Eref{eq:split_mbyr3},
we find that the RHS of ODEs \Eref{eq:tov_dxdmu} 
and \Eref{eq:tov_dladmu} remains finite and non-zero at the origin.
Knowing the behavior near the origin is required for the numerical 
integration, which will be discussed in \Sref{sec:impl_numeric}.

Finally, we discuss the solution
outside the star. Since $P=E=0$, \Eref{eq:tov_dladmu} implies that
$\lambda+\mu = \mathrm{const}$, and \Eref{eq:tov_dmdr} implies that
$m(r)=M$, where $M=m(R)$ is the gravitational mass of the NS and 
$R$ the proper circumferential surface radius.
From \Eref{eq:tov_m_of_lambda} we find $\lambda \rightarrow 0$ for
$r \rightarrow \infty$.
Using the standard gauge choice for the time, $\nu \rightarrow 0$
for $r \rightarrow \infty$,
we thus arrive at $\lambda=-\nu$ outside the star.
At the surface, \Eref{eq:tov_m_of_lambda} yields
\begin{align}
\nu(R) = -\lambda(R) &= \ln\left(\sqrt{1-\frac{2M}{R}}\right)
\end{align}
Further, $P(R)=0$ and therefore $H(R)=1$.
From \Eref{eq:tov_h_of_nu}, we find $\mu(R) = \ln (H(0))$. 
Since $\mu$ is our independent variable, the interval over which 
we need to integrate the ODE is known in terms of the central 
pseudo-enthalpy. Finally, 
we obtain $\nu(0) = \nu(R) - \mu(R) = \nu(R) - \ln(H(0))$.

\subsection{Tidal Deformability}
\label{sec:tidalode}
In the following, we describe a robust method for 
computing the tidal deformability of nonrotating NS 
in the limit of vanishing orbital frequency. 
We will start from the formulation derived in 
\cite{Hinderer:2008:533487,Hinderer:2008:533487:erratum,Postnikov:2010:024016} 
and then cast it into a form that allows numerical integration also
in the presence of phase transitions.

We make the standard assumption 
that the same barotropic EOS used to compute the unperturbed model 
also holds when the star is perturbed by a tidal field, in the limit 
of vanishing orbital frequency. 
It is worth pointing the physical implications behind this assumption.

First, the assumption is valid for the case of an isentropic barotopic
EOS, in particular for cold NS models.
For hot NS models on the other hand, this assumption will not 
hold in general. However, thermal effects can be safely neglected
for BNS systems near merger.

Second, NS matter also has another degree of freedom expressed
by the electron fraction. For the EOS describing a cold NS, it 
is given as function of density by the condition of $\beta$-equilibrium.
In the limit of very long orbital periods, $\beta$-equilibrium would be 
maintained, but such timescales will likely remain inaccessible to 
gravitational wave observations of BNS mergers. 
Here, we neglect the impact of any 
deviation of the electron fraction in the tidally perturbed star
from the one given by the EOS of the background model, and
do not try to estimate the corresponding error.

As detailed in \cite{Hinderer:2008:533487} (beware of a typo corrected in \cite{Hinderer:2008:533487:erratum}), 
the dimensionless tidal deformability $\Lambda$ is given by
\begin{align}\label{eq:tidal_def_lambda}
\Lambda &= \frac{2}{3 \beta^5}  k_2,  \qquad \text{where} & \beta &= \frac{M}{R} 
\end{align}
where the Love-number $k_2$ is given by
\begin{align}\label{eq:tidal_k2}
\begin{split}
k_2 &= \frac{8}{5} \beta^5 \left( 1 - 2 \beta \right)^2
\left( 2 + 2 \beta \left( y_s - 1 \right) - y_s \right)
\\ & \quad \times 
\left[ 
2 \beta  \left(6 - 3 y_s  + 3 \beta \left( 5 y_s - 8 \right) 
\right.
\right. \\ &  \quad \left.
\left.
       + 2 \beta^2 \left( 13 - 11 y_s + \beta \left(3 y_s - 2 \right) 
                          + 2 \beta^2 \left(1+y_s \right) \right) \right)
\right. \\ &  \quad \left.
       + 3 \left(1 - 2 \beta \right)^2 \left( 2 - y_s 
                 + 2 \beta \left( y_s -1 \right) \right) 
\right. \\ &\qquad \left.                 
                 \ln \left(1 - 2 \beta \right) 
\right]^{-1}
\end{split}
\end{align}
The quantity $y_s$ is the surface value $y_s \equiv y(R)$ of a 
radial function $y(r)$ that is the unique solution of the ODE
\begin{align}\label{eq:tidalode1}
r \frac{\mathrm{d}}{\mathrm{d}r} y 
&= - y^2 - y e^{2\lambda} \left( 1 + 4 \pi r^2 \left(P - E \right) \right) - r^2 Q 
\\ \label{eq:tidalode1_q}
\begin{split}
r^2 Q &= 4 \pi r^2 e^{2\lambda} \left( 5 E + 9 P + \frac{E+P}{c_s^2} \right)
\\& \quad
         - 6 e^{2\lambda} - 4 \left( r \frac{\mathrm{d}\nu}{\mathrm{d}r}  \right)^2 
\end{split}
\end{align}
which was derived in \cite{Postnikov:2010:024016} from the original formulation in 
\cite{Hinderer:2008:533487, Hinderer:2008:533487:erratum} (note our definitions
of $\lambda,\nu$ differ from \cite{Postnikov:2010:024016} by a factor 2).
The variable $y$ is defined as $y=r\mathcal{H}'/\mathcal{H}$, where 
$\mathcal{H}$ fulfills a second order ODE given in 
\cite{Hinderer:2008:533487} (note $\mathcal{H}$ is denoted $H$, not 
to be confused with our notion for the pseudo-enthalpy $H$).

The solution of $y$ is unique because the boundary conditions for 
$\mathcal{H}$ given in \cite{Hinderer:2008:533487} imply that $y(0)=2$. 
The behavior of $y$ for $r\to 0$ can be found using a Taylor expansion in $r$.
Together with \Eref{eq:split_mbyr3}, we find that
\begin{align}\label{eq:tidalode1_ybnd}
y &\approx 2 - \frac{4}{7} \pi \left(\frac{1}{3} E_c + 11 P_c 
       + \frac{E_c+P_c}{c_s^2} \right) r^2 
\end{align}
where $E_c$ and $P_c$ denote the central values.
We note that our result contradicts a similar expansion 
in terms of $\ln(H)$ provided in \cite{Postnikov:2010:024016}.   
For comparison, we use \Eref{eq:tov_dnudx} and \Eref{eq:tov_dhdr}
to express \Eref{eq:tidalode1_ybnd} as
\begin{align}
y & \approx 2 +
\frac{2}{7} 
\frac{E_c + 33 P_c + \frac{3}{c_s^2}(E_c+P_c)}{E_c + 3 P_c}
\ln\left(\frac{H}{H_c}\right)
\end{align}

It is important to note that the ODE coefficients in \Eref{eq:tidalode1_q} are 
degenerate at phase transitions (see also \cite{Postnikov:2010:024016}). 
The culprit is the term
$(E+P)/c_s^2$. We recall that phase transitions feature a range of mass density 
over which $P$ is constant and $c_s=0$. For hydrostatic solutions, the pressure 
is smooth across the phase transition while the density has a jump. 
Taken together, the ODE coefficient as function of radius acquires a 
delta-function component at a phase transition. Even if the EOS only 
has a plateau with nearly constant pressure, it results in a very sharp peak.
Integrating the original ODE numerically across a phase transition 
is impossible. One could compute the jump across a phase 
transition analytically, as done in 
\cite{Postnikov:2010:024016, Takatsy:2020:028501}. 
We do not follow this approach 
since phase transitions would then have to be described separately, 
thus complicating the EOS handling. 

\begin{figure}
\caption{The term $\rho/c_s^2$ as function of pseudo-enthalpy $H-1$, 
for various tabulated EOS models (described in \Sref{sec:eoscoll}) 
from nuclear physics. 
The pseudo-enthalpy be regarded as a proxy for the radial 
coordinate in NS models.
The inset highlights some of the sharp peaks, 
which are problematic for
the standard formulation of the tidal deformability ODE.
}\label{fig:phase_impact}
\includegraphics[width=\columnwidth]{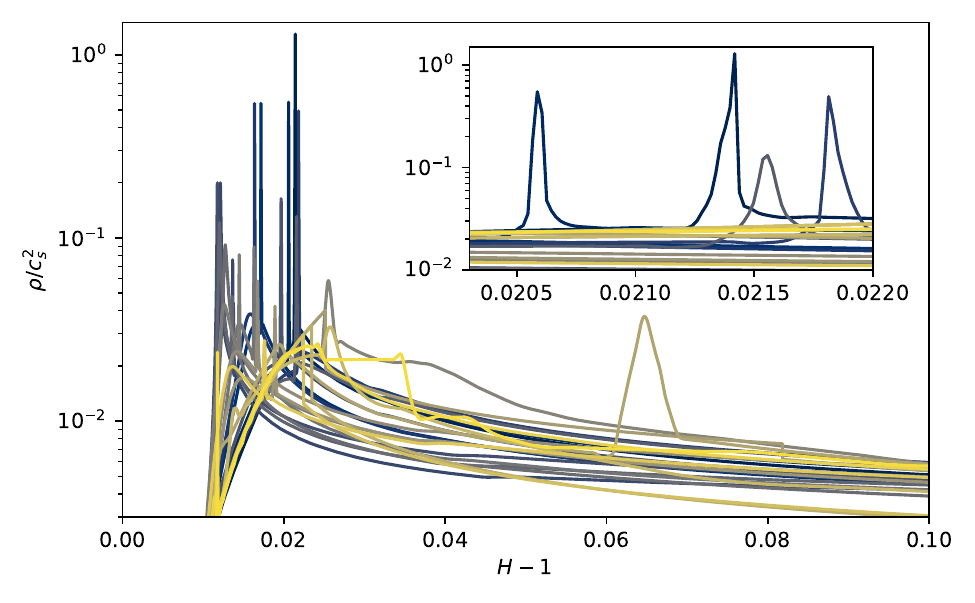}
\end{figure}

In practice, EOS can exhibit behaviors between an idealized phase 
transitions and merely sharp features, which also makes any 
semi-analytic approach ambiguous.
In \Fref{fig:phase_impact}, we show the problematic term $\rho/c_s^2$ 
for a selection of nuclear physics EOS models (see \Sref{sec:eoscoll}). 
Although these 
contain only very weak phase transitions, the resulting peaks 
are quite sharp and difficult to resolve.
In order to circumvent this problem altogether,
we will now transform the ODE into a form that
has regular coefficients at phase transitions and can be integrated 
using standard numerical methods.

We start by noting that the offensive term is closely related to the 
derivative of the density. 
\begin{align} \label{eq:tov_drhodnu}
\frac{E+P}{c_s^2} &= \frac{\rho h}{c_s^2}
= \rho h \frac{\mathrm{d}\ln \rho}{\mathrm{d}\ln H}
=  -h \frac{\mathrm{d}\rho}{\mathrm{d}\nu}
\end{align}
Above, we used \Eref{eq:eos_csnd_adiab_alt} and \Eref{eq:tov_dhdr}.

We can now simply cancel the degeneracy by using the density instead of the 
radius as independent variable.
\begin{align}
\frac{\mathrm{d}y}{\mathrm{d}\rho} 
&= \left(  \frac{\mathrm{d}\rho}{\mathrm{d}\nu}  \right)^{-1}
\left( \frac{\mathrm{d}y}{\mathrm{d}\nu} \right) \\
&= -\frac{h c_s^2}{E+P} \frac{\mathrm{d}y}{\mathrm{d}\nu}
= -\frac{c_s^2}{\rho} \frac{\mathrm{d}y}{\mathrm{d}\nu} \\
&= -\frac{c_s^2}{\rho} 
   \left(\frac{\mathrm{d}\nu}{\mathrm{d}x}  \right)^{-1}
   \left(\frac{\mathrm{d}y}{\mathrm{d}x}  \right)   
\end{align}
From \Eref{eq:tidalode1} and \Eref{eq:tidalode1_q}, we get
\begin{align}
\begin{split}
\frac{\mathrm{d}y}{\mathrm{d}x} 
&= 
\frac{1}{2x} \left[ 
-y^2 + \left( 6 - y \right) e^{2\lambda}
\right] 
+8x \left(\frac{\mathrm{d} \nu}{\mathrm{d}x}\right)^2
\\&\quad
-2\pi e^{2\lambda} \left[ \left( P-E \right) y  + 5E + 9P + \frac{\rho h}{c_s^2} \right]
\end{split}
\end{align}
Using the identity
\begin{align}
e^{2\lambda} - 1 &= 2 x e^{2\lambda} \frac{m}{r^3}
\end{align}
we can rewrite the first term as
\begin{align}
\begin{split}
\frac{1}{2x} &\left[ 
-y^2 + \left( 6 - y \right) e^{2\lambda}
\right] \\
&= - \frac{y-2}{2x} \left(y+3\right)
   + \left(  6 - y \right) e^{2\lambda} \frac{m}{r^3}
\end{split}
\end{align}
Collecting terms, we find
\begin{align}\label{eq:tidalode_dydrho}
\frac{\mathrm{d}y}{\mathrm{d}\rho} &=
\frac{4\pi h}{4 \pi P + \frac{m}{r^3}}
+ \frac{c_s^2}{\rho} A\\
\begin{split}
A &= 
\frac{2 }{4 \pi P + \frac{m}{r^3}}
\Big(
\frac{y-2}{2x}\left(y+3\right)e^{-2\lambda} + \left( y - 6 \right)\frac{m}{r^3} 
\Big.\\&\quad\Big.
+2 \pi \left( P-E \right) y + 2 \pi \left(5E +9P \right)
\Big)
\\ &\quad
- 4 x e^{2\lambda} \left(4 \pi P + \frac{m}{r^3} \right)
\end{split} \label{eq:tidalode_dydrho_A}
\end{align}
The coefficients of the new ODE for $y$ are finite across 
phase transitions. 
They are also finite at the center, as follows from 
\Eref{eq:tov_lt_by_r_lim} and \Eref{eq:tidalode1_ybnd}.
Even if a phase transition happens exactly at the center,
they remain finite. Although \Eref{eq:tidalode_dydrho_A} contains a 
term $(y-2)/x$ which in this case diverges (see
\Eref{eq:tidalode1_ybnd}) like $1/c_s^2$, the coefficient 
$c_s^2 A$ in \Eref{eq:tidalode_dydrho} remains finite.

At the surface, on the other hand, the term ${c_s^2}/{\rho}$ in
the new ODE is now problematic, because the limit of zero 
density is not finite for all EOS. For example, it diverges
for a polytropic EOS with $\Gamma<2$ (compare 
\Eref{eq:eos_poly_rho_of_h} and \Eref{eq:eos_poly_csnd_of_h}).

To solve the problem near the surface,
we derived yet another formulation, without sacrificing the
good behavior at phase transitions.
For this, we first change the dependent variable to
\begin{align}\label{eq:tidalode_def_yhat}
\hat{y}(\rho) &= y(\rho) - d(\rho) \\
d(\rho) &= \int_0^\rho \frac{4 \pi h}{4 \pi P + \frac{m}{r^3}} \mathrm{d} \rho'
\label{eq:tidalode_def_d_of_rho}
\end{align}
The term $mr^{-3}$ in the integrand can be written as function of density $\rho$
since the density is strictly decreasing with radius for solutions
of the TOV equations. The integrand is finite also at phase transitions,
where it simply exhibits a plateau. The integration can thus easily be 
carried out numerically.
The above subtraction eliminates the first term in \Eref{eq:tidalode_dydrho}, 
which becomes
\begin{align}
\frac{\mathrm{d}\hat{y}}{\mathrm{d}\rho} &= \frac{c_s^2}{\rho} A
= \frac{\mathrm{d}\ln(H)}{\mathrm{d}\rho} A
\end{align}
Finally, we change the independent variable to $\nu$,
writing
\begin{align}\label{eq:tidalode_dyhatdnu}
\frac{\mathrm{d}\hat{y}}{\mathrm{d}\nu} &= 
\frac{\mathrm{d}\hat{y}}{\mathrm{d}\rho}
\frac{\mathrm{d}\rho}{\mathrm{d}\nu}
=
-\frac{\mathrm{d}\hat{y}}{\mathrm{d}\rho}
\frac{\mathrm{d}\rho}{\mathrm{d}\ln(H)}
=
- A
\end{align}
The RHS given by \Eref{eq:tidalode_dydrho_A} is completely
unproblematic both at phase transitions and at the surface.
The discontinuity in $y$ is now contained entirely in the variable $d$.
Although $d(\rho)$ is smooth, $\rho(\nu)$ and therefore 
$d(\rho(\nu))$ have a jump at a phase transition.

As a crosscheck, we can recover the analytical contribution of a 
phase transition discussed in \cite{Postnikov:2010:024016, Takatsy:2020:028501}.
Denoting the density range of the phase transition by $[\rho_-, \rho_+]$,
the radial location by $r_T$, and the constant enthalpy and pressure 
across the transition by $h_T$ and $P_T$, respectively, we find
\begin{align}
d(\rho_+) - d(\rho_-) &= 
\left(\rho_+ - \rho_- \right) 
\frac{4 \pi h_T}{4 \pi P_T + \frac{m(r_T)}{r_T^3}} 
\end{align}
since the integrand in \Eref{eq:tidalode_def_d_of_rho} is constant across
the phase transition. Using $\hat{y}_+=\hat{y}_-$, 
a straightforward computation yields
\begin{align}
\Delta y &\equiv y_+ - y_- = 
\frac{4 \pi \left( E_+ - E_-\right)}{4 \pi P_T + \frac{m(r_T)}{r_T^3}} 
\end{align}
which agrees with \cite{Takatsy:2020:028501}.
A similar result in \cite{Han:2019:083014} seems to be intended to describe phase 
transitions near the surface and agrees in the limit of small $P_T$.

Normally, the above ODE behaves well also at the center. Only in the 
rare case where the central density coincides with a phase transition,
the coefficient $A$ diverges like $1/c_s^2$ (compare \Eref{eq:tidalode1_ybnd}).
To sidestep this remaining problem, we use the formulation 
given by \Eref{eq:tidalode_dydrho} to integrate $y$ up to some radius inside the star,
and then switch to the formulation given by \Eref{eq:tidalode_dyhatdnu} to integrate
$\hat{y}$ up to the surface.

\subsection{Moment of Inertia}
\label{sec:minertia}
The moment of inertia of uniformly rotating NS in the slow rotation
limit was derived in \cite{Hartle:1967}. This involves the solution of 
an ODE for a 
function $\bar{\omega}$ related to frame dragging, with coefficients 
determined by the TOV solution describing the nonrotating NS.
In detail,
\begin{align}\label{eq:minert_hartle_ode}
\frac{\mathrm{d}^2\bar{\omega}}{\mathrm{d}r}
&= 4 \pi \rho h e^{2\lambda} 
     \left( r \frac{\mathrm{d}\bar{\omega}}{\mathrm{d}r} + 4 \bar{\omega} \right)
     - \frac{4}{r} \frac{\mathrm{d}\bar{\omega}}{\mathrm{d}r}
\end{align}
Compared to the form in \cite{Hartle:1967}, we inserted 
\Eref{eq:tov_dladr} and \Eref{eq:tov_dnudr}.
To solve the ODE,
we first transform it into an equivalent first order ODE 
using $x$ as independent variable
\begin{align} \label{eq:minert_ode}
\frac{\mathrm{d}\bar{\omega}}{\mathrm{d}x}
&= \frac{\bar{\omega}_1}{x} \\\label{eq:minert_ode1}
\frac{\mathrm{d}\bar{\omega}_1}{\mathrm{d}x} 
&= \left( 2\pi x \rho h e^{2\lambda} - \frac{3}{2} \right) \frac{\bar{\omega}_1}{x}
   + 4 \pi \rho h e^{2\lambda}  \bar{\omega}
\end{align}
Another variable transform
leads to a form that can be integrated simultaneously with the TOV 
equations
\begin{align}
\frac{\mathrm{d}\bar{\omega}}{\mathrm{d}\mu}
&= \left( \frac{\mathrm{d}x}{\mathrm{d}\mu} \right)
   \frac{\mathrm{d}\bar{\omega}}{\mathrm{d}x} 
,&
\frac{\mathrm{d}\bar{\omega}_1}{\mathrm{d}\mu} 
&= \left( \frac{\mathrm{d}x}{\mathrm{d}\mu} \right) 
   \frac{\mathrm{d}\bar{\omega}_1}{\mathrm{d}x}
\end{align}
where ${\mathrm{d}x}/{\mathrm{d}\mu}$ is given by 
\Eref{eq:tov_dxdmu}.

The coefficients of ODE \Eref{eq:minert_hartle_ode} are degenerate at 
the center, and the solution has just one degree of freedom given by 
$\bar{\omega}(0)$. Using a Taylor expansion, one can show that 
the solution for $\bar{\omega}_1$ in the limit $r\rightarrow 0$ 
is
\begin{align}\label{eq:minert_origin_limit}
\bar{\omega}_1 &\rightarrow 0, &
\frac{\bar{\omega}_1}{x} &\rightarrow \frac{8}{5} \pi \rho h\bar{\omega}
\end{align}

As shown in \cite{Hartle:1967}, the function $\bar{\omega}$ outside the star  
is related to the angular momentum $J$ of the NS and its angular velocity 
as seen from infinity, $\Omega$, by
\begin{align}
\bar{\omega}(r) &= \Omega - \frac{2J}{r^3} \qquad r \ge  R 
\end{align}
As a consequence,
\begin{align}
J &= \left.\frac{r^4}{6}\frac{\mathrm{d}\bar{\omega}}{\mathrm{d}r} \right|_{r=R}
\end{align}
which is unambiguous since
$\bar{\omega}$ is twice differentiable also across the NS surface.
Combining the equations above, we compute the moment of inertia 
using the expression
\begin{align}
I &\equiv \frac{J}{\Omega}
= 
\left. \frac{r^3}{3\frac{\bar{\omega}}{\bar{\omega}_1}         
           + 2 } \right|_{r=R} 
\end{align}
Compared to \cite{Hartle:1967}, this avoids another integration step. 
\subsection{Baryonic Mass, Binding Energy, Volume}
\label{sec:barymass}
The baryonic mass within a radius $r$ is given by the integral
\begin{align} \label{eq:miscns_mb_int}
M_b(r) &= 
\int_0^r 4\pi  {r'}^2 \sqrt{g_{rr}(r')} \rho(r')  \mathrm{d}r' \\
&= \int_0^r 4\pi  {r'}^2 e^{\lambda(r')} \rho(r') \mathrm{d}r'
\end{align}
where we use the coordinates \Eref{eq:tov_line_element}.

The total baryonic mass of a NS, $M_b = M_b(R)$, is important 
in the context of 
binary neutron star mergers. The law of baryon number conservation
implies that the total baryonic mass of the system after merger
is given by the sum of the baryonic mass of original coalescing 
NS. This can be used to relate the stability of the remnant to 
the EOS.

The binding energy of a NS is usually defined as $E_b = M_b - M$.
We note that this definition is a slight misnomer since $E_b$ 
is not exactly the energy difference between
the ADM energy of a NS and a spacetime where the same amount of 
matter is infinitely dispersed. That would only be the case
if each baryon contributes an energy given by the arbitrary 
constant $m_b$ for the dispersed state, whereas the actual 
value would depend on the nuclear composition of the matter.

Another quantity we compute is the proper volume $V(r)$ 
enclosed within spherical surfaces of radius $r$,
which is given by
\begin{align}\label{eq:miscns_vol_int}
V(r) 
&= \int_0^r 4\pi  {r'}^2 e^{\lambda(r')} \mathrm{d}r'
\end{align}
The proper volume of the NS is then given by $V(R)$. 
We are unaware of existing results providing $V(r)$ outside 
the star, and derived the analytic expression below using 
straightforward integration.
\begin{align}\label{eq:miscns_vol_int_outside}
\begin{split}
V(r)  
&= V(R) + 
\frac{4}{6} \pi  \left[
15 M^3 \ln\left(r'\sqrt{1-\frac{2M}{r'}} + r' - M \right)
\right.\\&\quad\left.
+ r'\sqrt{1-\frac{2M}{r'}} \left(
M\left(15 M + 5 r'\right) + 2 {r'}^2
\right)
\right]_R^r
\end{split}
\end{align}

It is convenient to compute $E_b(r) = M_b(r) - M(r)$ 
and $V(r)$ while solving the TOV ODE instead of using a 
separate integration step.
For this purpose, we can cast \Eref{eq:miscns_mb_int} and 
\Eref{eq:miscns_vol_int} into the differential equations
\begin{align}
\frac{\mathrm{d}}{\mathrm{d}x} \left(\frac{V(r)}{r} \right)
&= 2 \pi e^\lambda - \frac{V(r)}{2r^3} 
\label{eq:miscns_pvol_ode} \\
\frac{\mathrm{d}}{\mathrm{d}x} \left(\frac{E_b(r)}{r} \right)
&= 2 \pi \rho \left( e^\lambda - 1 - \epsilon \right) 
   - \frac{E_b(r)}{2 r ^3}
\label{eq:miscns_ebnd_ode}
\end{align}
For $r\rightarrow 0$, we find
\begin{align}\label{eq:miscns_ebnd_pvol_bndry}
V(r)
&\rightarrow \frac{4}{3}\pi r^3, &
E_b(r)
&\rightarrow -\frac{4}{3}\pi \rho \epsilon r^3
\end{align}
The advantage of this form is that the quantities $V/r$ and $E_b/r$ 
grow linearly with $x$ near the origin and the RHS of 
\Eref{eq:miscns_pvol_ode} and \Eref{eq:miscns_ebnd_ode} remains 
finite.

\subsection{Bulk measures}
\label{sec:bulkmeasures}

The proper volume is rarely used, which is surprising since it is 
a fundamental geometric property of any body, not restricted to 
spherical symmetry. This 
was utilized in the context of BNS merger remnants in \cite{Kastaun:2016}.
To define measures applicable to any spacetime, one can use proper 
volume $V(S_\rho)$ and baryonic mass $M_b(S_\rho)$ enclosed within 
isodensity surfaces $S_\rho$. One can also define 
an volumetric radius $R_V(S)$ as the radius of the Euclidean sphere 
with equal volume. This allows to define a compactness measure
$C_V(S) =  M_b(S) / R_V(S)$. Although merger remnants lack a clearly 
defined surface, there is always one surface $S_\text{blk}$ of maximum 
compactness $C_V$, denoted as the bulk surface in \cite{Kastaun:2016}. 
This gives rise
to definitions of bulk proper volume $V_\text{blk}=V(S_\text{blk})$
and bulk baryonic mass $M_\text{blk} = M_b(S_\text{blk})$. 

The same measures can easily be applied to a TOV solution, where
the isodensity surfaces $S_\rho$ are just spheres $S_r$. 
For a static NS, the formula given in \cite{Kastaun:2016} to 
determine the bulk surface simplifies to
\begin{align}
\rho(r_\text{blk}) &= \frac{M_b(r_\text{blk})}{3 V(\rho_\text{blk})}
\end{align}

The bulk measures above are used in \cite{Kastaun:2016} to compare the cores 
of TOV and merger remnants in a well-defined way. In detail,
one can search for the intersection between the proper mass-volume 
relation of the remnants isodensity surfaces and the bulk 
mass-volume relation for the sequence of TOV solutions. If it exists,
the corresponding TOV solution is called TOV core equivalent,
because its bulk has a similar radial mass distribution than
the merger remnant core.
 
\subsection{Stable Circular Orbits}

It is a well known fact that circular orbits of massive test 
particles in the Schwarzschild metric are unstable at radii 
smaller than the innermost stable circular orbit (ISCO) located 
at $R_\text{ISCO} = 6M$. Since the metric outside a spherical
NS is the Schwarzschild metric, all orbits outside the 
NS surface are given by the latter. Therefore, all NS
with compactness $M/R > 1/6$ have an ISCO. 
Realistic NS models do indeed have an ISCO above some mass
that depends on the EOS.

We could not find any reference regarding the stability of
geodesics inside NS.
This seems a somewhat academic question, but might be relevant 
when studying accumulation of particle dark matter inside NS.
Below, we provide a short discussion of subsurface orbits.
Interestingly, it turns out that NS with an ISCO also have an 
outermost stable internal circular orbit (OSICO in the following). 

We start by setting up the usual geodesic invariants. 
For the line element \Eref{eq:tov_line_element}, both $\partial_t$ and $\partial_\phi$
are Killing vectors. For a geodesic curve $x^\mu(\tau)$,
we find the invariants
\begin{align}
L &= \dot{x}_\phi = r^2 \dot{x}^\phi \\
E &= -\dot{x}_t = \dot{x}^t e^{2\nu}
\end{align}
where we have assumed that the curve is parametrized 
such that $\dot{x}^\mu \dot{x}^\nu g_{\mu\nu} = -1$.
From the above, it immediately follows that
\begin{align}
\frac{1}{2} E^2 &=
V(r) + \frac{1}{2} e^{2\left(\lambda+\nu\right)} \left(\dot{x}^r\right)^2\\
V(r) &\equiv \frac{1}{2} e^{2\nu} \left(1 + \frac{L^2}{r^2} \right)
\end{align}
For a given angular momentum $L$, the minima of the effective 
potential $V(r)$ correspond to stable circular orbits, and the 
maxima to unstable circular orbits. It is trivial to compute
\begin{align}
V'(r) &= \frac{e^{2\nu}}{r}\left( r \nu' 
          - \frac{L^2}{r^2} \left( 1 - r \nu' \right) \right),
\end{align}
The condition $V'=0$ yields the angular momentum $L^c(r)$ 
for circular orbits at radius $r$
\begin{align}
L^c(r) &= r \sqrt{\frac{r\nu'}{1-r\nu'}}
\end{align}
Note that there is no circular orbit if $r\nu' > 1$.
This happens for the Schwarzschild metric below 
$r<3M$ (the location of the photonsphere).
For simplicity, we will ignore the potential corner case 
of NS with $R \le 3M$ or NS that violate $r\nu' < 1$ 
anywhere within the interior.
We can thus assume the existence of circular orbits at 
any radius inside a NS.

\begin{figure}
\caption{Angular momentum of circular orbits for radii 
outside and inside NS employing the \texttt{APR4} EOS.
The masses shown range from the lowest mass for which there
is an ISCO to the maximum mass. The circles mark the surface 
location. 
The solid curves correspond to the stable orbits, the 
dotted curves to unstable ones, and the OSICOs are marked
by crosses.
For comparison, we show the angular momentum for orbits
around Schwarzschild BH (red), and the location of the 
ISCO (plus symbol).
}\label{fig:orbits}
\includegraphics[width=\columnwidth]{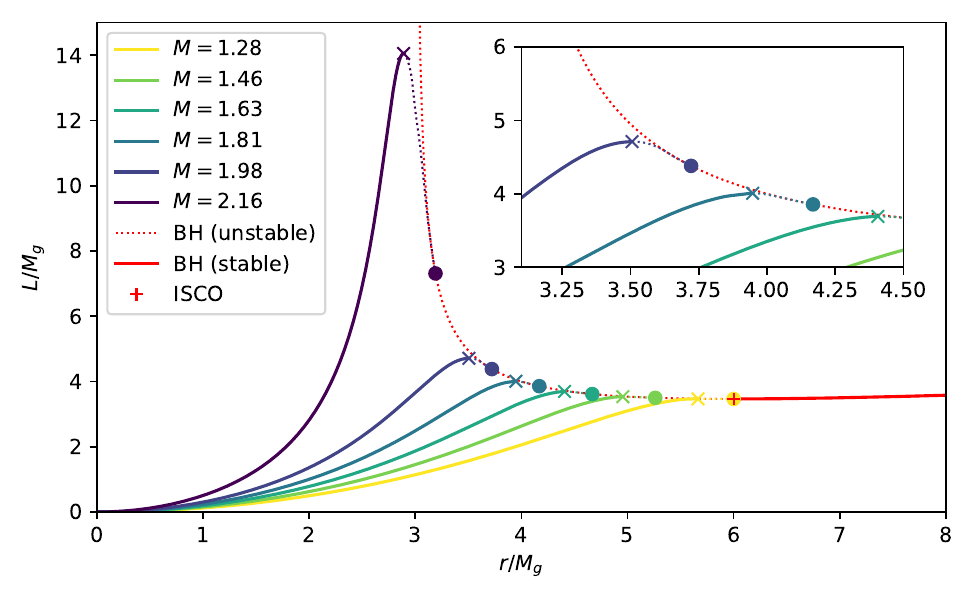}
\end{figure}

Next, we discuss the stability of the internal circular orbits.
For a given $L$, the roots of $L^c(r) - L$ are the extrema 
of $V(r)$. Starting from the origin, the first root must 
correspond to a minimum
since $V(r) \to \infty$ for $r \to 0$. If there is a second root,
it corresponds to a maximum (for the latter case,
there must be a third root since $L^c\to\infty$ for $r \to \infty$).
Therefore, circular orbits in the interval $(0, r_O)$ are stable,
where $r_O$ is the location of the first maximum
of $L^c$ within the star. Further, $L^c$ cannot have a 
maximum at the surface, where $\mathrm{d}L^c/\mathrm{d}r < 0$
(still assuming $R> 3M$).
The behavior of $L^c$ is illustrated in \Fref{fig:orbits} for
one EOS and different masses.
To conclude, we find that NS with $3M < R < 6M$ feature an ISCO
outside the star and an OSICO located strictly below the surface.

We now turn to the orbital angular velocity of test 
particles on circular orbits. A general expression valid 
for axisymmetric stationary 
spacetimes can be found, e.g., in \cite{Laarakkers:1999:282}. 
Using \Eref{eq:tov_line_element} and \Eref{eq:tov_dnudr},
we can specialize to the spherical NS case to obtain
\begin{align}\label{eq:omega_of_r_full}
\Omega(r) &= \sqrt{\frac{-g_{tt,r}}{g_{\phi\phi,r}}}
 = e^{\nu+\lambda} \sqrt{4\pi P +\frac{m}{r^3}}
\end{align}
Outside the star, this simplifies to
\begin{align}
\Omega(r) &= \sqrt{\frac{M}{r^3}}
\end{align}
Curiously, the orbital velocity assumes a nonzero value at 
the center. We are unaware of a corresponding formula in the 
literature, but it is trivial to derive. Using \Eref{eq:tov_lt_by_r_lim} and
\Eref{eq:omega_of_r_full}, we obtain
\begin{align}\label{eq:omega_orb_center}
\Omega(0) &= \frac{1}{H_c}\sqrt{\frac{4}{3}\pi\left(E_c + 3P_c \right)}
          \sqrt{1 - \frac{2M}{R}}
\end{align}
This provides a rough scale for the rotational velocity at which one 
should expect strong deformation of the core. For example, numerical 
merger simulations have found differential rotation profiles where
the core is rotating much slower than the central orbital velocity,
and radial mass distributions very similar to a nonrotating NS
\cite{Kastaun:2015:064027,Kastaun:2016,Kastaun:2017}.

\section{Implementation}
\label{sec:impl}
In the following, we provide important technical details for
the numerical implementation of the equations given in the 
previous section.

\subsection{Avoiding numerical problems}
\label{sec:impl_numeric}
The use of finite precision arithmetic can lead to severe
accuracy problems. There are several 
places in our implementation where cancellation errors
would lead to a catastrophic loss of accuracy
with a naive implementation of our analytic formulas.

To avoid such a problem, we generally represent the 
pseudo-enthalpy $H$ in terms of $H_1 \equiv H-1$.
We recall that $H(\rho) \rightarrow 1$ for $\rho \rightarrow 0$. 
At low density, computing $\rho(H)$ or $P(H)$ would
be inaccurate when using $H$ directly.
Further, when evaluating the function $e^a-1$ for small arguments,
we use a numerical implementation, 
denoted here as $\mathrm{EXPM1}(a)$, 
that is accurate for $a\rightarrow 0$,
such that $\mathrm{EXPM1}(a) \approx a$. 
Evaluating $e^a-1$ directly with finite precision arithmetic 
would result in catastrophic loss of accuracy.

Another type of problem is the correct treatment of 
ODE coefficients at boundaries. 
The coefficients of the TOV ODE in our formulation are finite 
and continuous when evaluated along a solution of the ODE.
At $r=0$ the expressions \Eref{eq:tov_dxdmu}, \Eref{eq:tov_dladmu} 
are, however, only defined as 
a mathematical limit and cannot be evaluated numerically.
Instead one has to use the expression for the limit.
Further, the partial derivatives of the coefficients with 
respect to $\lambda$ and $x$ diverge at the origin. This 
leads to a reduction of the convergence order for the 
first ODE integration step at the origin. 
When using an RK4/5 integration scheme with adaptive step 
size control, the impact on the global accuracy 
that can be achieved with given costs is small. 
However, we find that the maximum convergence order that can 
be reached with a fixed step size is limited to second order.

In our implementation, we evaluate the problematic 
term $m /r^{3}$ as follows
\begin{align}\label{eq:tov_m_of_lambda_robust}
\frac{m}{r^3} &= 
\begin{cases}
-\frac{1}{2x} \mathrm{EXPM1}\left(-2\lambda\right) & \mu > \Delta  \\
\kappa_1(E_0, E(\mu), x) \sigma(-2\lambda) & \mu \le \Delta
\end{cases}
\end{align}
Above, $\Delta$ denotes the stepsize when using fixed-step 
integrator, and $\Delta =0$ when using an adaptive step size 
control.
The error caused by using the approximation $\kappa_1$ 
defined in \Eref{eq:def_kappa1} is of order $\mathcal{O}(\Delta^2)$.
This limits the maximum convergence order for the ODE solution
with fixed step size to 3rd order (one order higher than the error
of the local RHS coefficient because it is integrated over the first 
step only). Compared to the order reduction present when 
using the exact formula, using the approximation thus increases 
the possible convergence order by one.

The integration of the moment of inertia ODE is mostly
unproblematic, except for the term $\bar{\omega}_1/x$ 
in \Eref{eq:minert_ode}. For $x=0$, we use the analytic limit given 
by \Eref{eq:minert_origin_limit}. Small $x$ on the other hand are 
not a problem because $\bar{\omega}_1$ grows linearly from 
$\bar{\omega}_1(0)=0$.

When integrating \Eref{eq:tidalode_dydrho}, there is one term which 
needs special care when evaluated numerically in the limit 
$r\rightarrow 0$, although analytically it is finite
\begin{align}
\begin{split}
\lim_{x\rightarrow 0} & \frac{c_s^2}{\rho} \frac{y-2}{x} = \\
& -\frac{4}{7}\pi \left( h + 
\left(11 h - \frac{32}{3} \left(1+\epsilon\right)\right) c_s^2
              \right)
\end{split}
\end{align}
First, we use $y-2$ as dependent ODE variable instead of $y$.
Otherwise the term $y-2$ would suffer catastrophic loss of accuracy 
for $x\rightarrow 0$ (where $y\rightarrow 2$) due to cancellation 
errors. Second, when evaluating the term above at $x=0$, we use the 
analytic limit.

To compute proper volume and baryonic mass of a NS, we use the ODEs
given by \Eref{eq:miscns_pvol_ode} and \Eref{eq:miscns_ebnd_ode},
using \Eref{eq:miscns_ebnd_pvol_bndry} when $x=0$.
The choice of dependent variables is more suitable for adaptive 
ODE solvers since they grow linearly with $x$, in contrast to the 
straightforward ODEs for $V(x)$ and $M_b(x)$. Using the binding 
energy instead $M_b$ avoids
cancellation errors in the Newtonian limit (which is not a problem for 
NSs, but might be useful as a test case).

A third type of potential problems is the behavior of EOS and 
ODEs for NS properties at phase transitions. The problems 
with degenerate ODE coefficients are already taken care of 
by our analytical formulation, which also largely alleviates
problems caused by inaccurate numerical representations of 
an EOS across a phase transition. Below,
we provide the technical details of the corresponding 
implementation.

In order to compute the function $d(\rho)$ defined in 
\Eref{eq:tidalode_def_d_of_rho}, we first perform a numerical 
integration based on the sample
points obtained during the numerical solution of the TOV ODE.
This step is not problematic since the integrand is finite and 
continuous everywhere, including the origin, the surface, and phase
transitions. However, we found it necessary to use a 3rd-order 
accurate integration method (based on local quadratic interpolation)
in order to not restrict the overall convergence order.
The result of the integration is then used to construct
a monotonic interpolation spline for $d(\rho)$.
We also construct interpolation splines for $\lambda$ and $m/r^3$ as 
functions of $\rho$. It is important to interpolate $m/r^3$ instead of 
$m$ and $r$, in order to avoid amplification of interpolation errors 
for $r\rightarrow 0$.
Using those splines and the EOS allow to evaluate 
\Eref{eq:tidalode_dydrho} and \Eref{eq:tidalode_dydrho_A}.

To compute the tidal deformability, we use \Eref{eq:tidalode_dydrho}
to integrate $y-2$ from the central density to some lower density,
compute $\hat{y}$ using \Eref{eq:tidalode_def_yhat},
and then use \Eref{eq:tidalode_dyhatdnu} to integrate $\hat{y}$
to the surface. The exact matching point is not important, 
as we shall see. By default, our implementation uses 
$\log(H_\text{match}) = 0.1 \log(H_\text{center})$.

We remark that \Eref{eq:tidalode_dydrho} requires that the EOS implementation
can accurately compute $c_s(\rho)$ across a phase transition,
whereas \Eref{eq:tidalode_dyhatdnu} does not even use $c_s$. As long as 
the phase transition is below the matching density used for a given NS model,
one can compute the tidal deformability accurately even if the EOS 
implementation does not resolve the sound speed across the phase transition.

It is also worth pointing out that it is much more difficult to
numerically represent $c_s$ as function of $H$ than to represent
it as function of $\rho$, in particular when using interpolation 
of tabulated data. 
The reason is that $c_s(\rho)$
merely stays zero over an interval, while for $c_s(H)$ the entire phase 
transition is represented by a single point. Integrating the standard
formulation \Eref{eq:tidalode1} of the ODE across a phase transition 
would not just require some specialized ODE solver, but also require a
special EOS implementation able to represent extremely sharp features
in the soundspeed $c_s(H)$.

Another formula affected by large cancellation errors is \Eref{eq:tidal_k2},
as was pointed out in \cite{Postnikov:2010:024016}.
When applied to stars with compactness much lower than typical NS, such as 
white dwarfs, several orders of the terms polynomial in $\beta$ cancel
with the logarithmic term, as can be seen by Taylor-expanding it.
The problem becomes manifest when testing our initial implementation on 
unrealistic polytropic EOS that result in stellar models with 
compactness $\beta<0.01$. 

In \cite{Postnikov:2010:024016}, one can find
an approximation based on Taylor-expansion that solves the cancellation 
problem. However, by numerical comparison to the exact formula in the 
regime where both approximation and exact formula are accurate, we find 
that the expansion contains several faults. We therefore re-derived 
the expansion in $\beta$. For this, we approximate the love number as 
a polynomial as follows
\begin{align}\label{eq:k2_taylor}
\hat{k}_2(\beta, y) &=
  \left(1 - 2 \beta\right)^{2} 
  \sum_{l=0}^5 \beta^{l} p_{l}(y)
\end{align}
Expanding the master equation \Eref{eq:tidal_k2} in powers of $\beta$,
we obtain the coefficients
\begin{align}  
p_0(y) &= - \frac{y - 2}{2 \left(y + 3\right)} \\
p_1(y) &= \frac{y^{2} + 6 y - 6}{2 \left(y + 3\right)^{2}} \\
p_2(y) &= \frac{y^{3} + 34 y^{2} - 8 y + 12}{14 \left(y + 3\right)^{3}} \\
p_3(y) &= \frac{y^{4} + 62 y^{3} + 84 y^{2} + 48 y + 36}{14 \left(y + 3\right)^{4}} \\
p_4(y) &= \frac{5}{294} \left(5 \, y^{5} + 490 \, y^{4} + 1472 \, y^{3} + 1884 \, y^{2} 
\right. \\ &\qquad \left.       
        + 1476 \, y + 648\right) \left(y + 3\right)^{-5} \\
p_5(y) &= \frac{1}{294 } \left(33 y^{6} + 4694 \, y^{5} + 22100 \, y^{4} 
\right. \\ & \qquad 
        + 46440 \, y^{3}  + 57240 \, y^{2}  + 42984 \, y 
 \\ & \qquad \left.         
        + 15552\right)\left(y + 3\right)^{-6}
\end{align}
We verified numerically that this expansion converges to the exact formula
with error ${\sim}\beta^6$ until the numerical cancellation errors start to dominate.
For the final numerical implementation, we use 
\begin{align}\label{eq:k2_robust_impl}
k_2 &= 
\begin{cases}
k_2(\beta, y) & \beta>\beta_\text{thr} \\
\hat{k}_2(\beta, y) 
        + \left(\frac{\beta}{\beta_\text{thr}}\right)^6
        \left.\left(k_2- \hat{k}_2\right)\right|_{(\beta_\text{thr}, y)}
        & \beta \le \beta_\text{thr}
\end{cases}
\end{align}
Here 
$\beta_\text{thr}=0.05$ is a threshold value chosen based on plots
of $k_2$ comparing exact expression and approximation formula. 
Further,
$k_2(\beta, y)$ and $\hat{k}_2(\beta, y)$ 
are the exact and approximate expressions given by 
\Eref{eq:tidal_k2} and \Eref{eq:k2_taylor}.
The addition of the sixth-order term proportional to the difference
between the two at the threshold value reduces the error further, 
although it remains of order $\beta^6$.

\subsection{Interpolating EOS}

For EOS provided only at sample points, some form of interpolation is 
required. The interpolation method needs to be monotonic in order to 
prevent overshoots that violate \Eref{eq:eos_cs2_limits}. It is also
desirable that the interpolated functions are differentiable because
otherwise the convergence order of any ODE solver drops as soon as 
the ODE step becomes comparable to the EOS sampling resolution.
The sample points should cover many orders of magnitude in density 
and therefore is is best to perform the interpolation in logarithmic 
space. Finally, the numerical costs of interpolation should ideally not increase
with number of sample points. To satisfy all those requirements,
we use a monotonic cubic spline interpolation of
$\ln P(\ln H_1)$, $\epsilon(\ln H_1)$, $\ln \rho(\ln H_1)$, 
$\ln H_1(\ln \rho)$, and $c_s(\ln \rho)$.
The samples are spaced regularly in $\ln(\rho)$ or $\ln(H_1)$.
We interpolate $c_s$ in terms of $\rho$ instead of $H_1$ because it is 
impossible to resolve a phase transition for the latter case since it 
corresponds to a single point in $H$ but a range in $\rho$.
From the above interpolating functions, we consistently compute
$P(\rho) = P(\ln H_1(\ln \rho))$ and $c_s(H_1) = c_s(\ln \rho(\ln H_1))$.

EOS samples are typically not provided with the particular spacing 
described above. We therefore perform another interpolation step
to create the required regularly spaced samples from the available 
ones. This is done as described above, only that we use a slower  
variant of the monotonic spline interpolation that allows non-regular
spacing.

Often, EOS samples are only provided above some low density cutoff.
Further, it is generally wasteful to use a large number of sample points
for densities too low to have any impact on NS properties.
Therefore, we restrict the range for interpolation above a suitably 
chosen cutoff density, which may be equal or larger than the 
lowest available sample point.
Below the interpolated range, we attach a polytropic EOS using the 
matching conditions \Eref{eq:eos_match_poly}.

When solving TOV equations, the EOS is required down to zero density 
in order to reach the NS surface.
Ideally, the sampled region extends to low enough densities such 
that the exact choice
for the extrapolation to low densities does not matter.
We will investigate this in \Sref{sec:tests_eos}.
In any case, the prescription provides a clearly defined EOS.

We note that it is much easier and more consistent to first extend 
an incomplete EOS to zero density than to add various technical 
workarounds during the computation of NS properties. 
Using workarounds in the TOV solver is either inconsistent or 
equivalent to some extrapolation of the EOS, with the disadvantage 
of not being made explicitly.

\subsection{Testing ODE Solutions}
\label{sec:tests}

In order to assess the accuracy of the 
numerical ODE solutions, we first study the 
convergence behavior for the simple case of
integrating the ODEs using fixed step size, employing a RK4/5
integrator.
We use twice as many points for the tidal ODE than for the
TOV ODE. We recall that TOV and tidal ODEs use different
independent variables and therefore the integrations steps 
for each ODE are not constant in terms of the 
independent variable of the other ODE. 
The factor of two was determined by roughly optimizing
computational costs at a given accuracy.

We are varying the step size for the TOV solution from 
10 to 10000 points within the NS. 
As an estimate for the ODE integration error, we compute the 
residuals of NS properties with respect to an even higher resolution of 
100000 points. 

We perform this test for many different EOS, each time 
for a NS model with $M_g = 1.4 \,M_\odot$ and for the maximum mass
model. We use EOS from three different categories.
The first group consists of 24 nuclear physics EOS represented as
monotonic spline interpolation of tabulated data. Those
EOS are described in \Sref{sec:eoscoll}.
The second group consists of piecewise polytropic approximations
of the MPA1 and MS1 EOS.
The last group contains polytropic EOS with polytropic indices
in the range $1\ldots 2.5$. The polytropic constant is chosen 
such that the maximum mass is $2.2\,M_\odot$. 

\begin{figure*}
\caption{Accuracy of solution versus resolution when using 
ODE solvers with fixed stepsize.
The panels show the errors of different quantities: gravitational mass $M_g$,
baryonic mass $M_\mathrm{B}$, proper volume $V$, circumferential radius $R$,
moment of inertia $I$, and tidal deformability $\Lambda$. 
The different curves show results for many spline-based EOS examples, 
as well as a family of polytropic EOS, and several piecewise polytropic ones.
The solid lines show models with $M_g=1.4\, M_\odot$ and the dashed lines 
show the maximum mass models (where the maximum is computed only once and 
the central density is kept fixed for all resolutions).
The thick green straight lines show the resulting empirical estimate for 
the resolution required to achieve given accuracies.
}\label{fig:accuracy_fixstep}
\includegraphics[width=\textwidth]{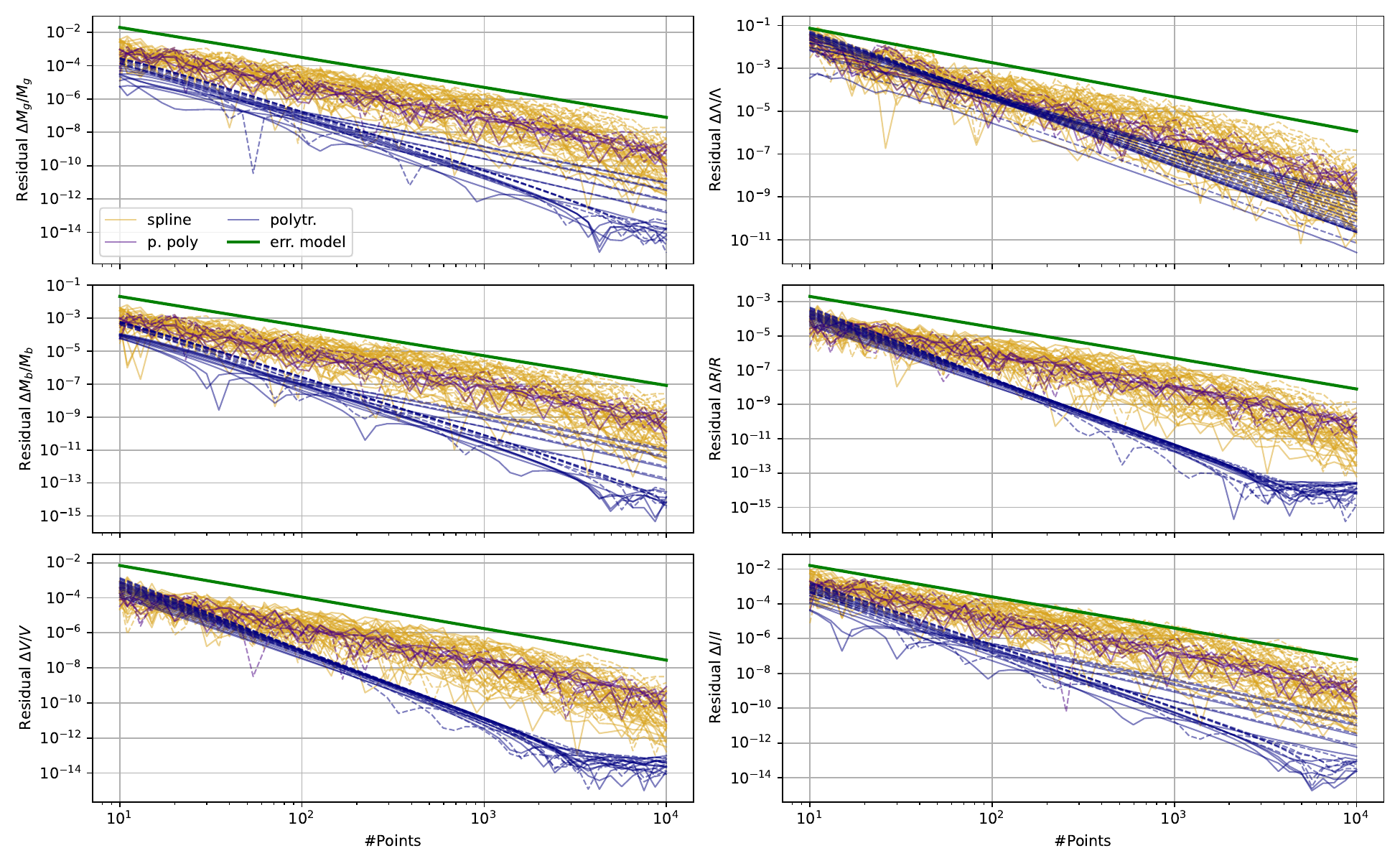}
\end{figure*}

\Fref{fig:accuracy_fixstep} shows the residuals for 
gravitational mass, baryonic mass, circumferential radius, 
proper volume, moment of inertia, and the tidal deformability.
We find that the solutions converge with increasing resolution,
but the convergence rate varies between the EOS,
and the absolute error differs strongly.
Not surprising, the errors are, on average, lowest for the 
analytic polytropic EOS. The piecewise polytropic EOS show
slower convergence, which we attribute to the fact that
they are not differentiable across the segment boundaries.
For the spline-based EOS, we observe a somewhat more noisy
convergence behavior. This might 
be caused by the interplay between the locations of ODE 
integration steps and EOS interpolation sample points.
Further, some of those EOS exhibit relatively sharp features 
in the sound speed (see \Fref{fig:phase_impact}).
Although those are not problematic, thanks to our choices of 
analytic formulations, they can still cause a non-constant 
convergence rate.

We note that for some of the polytropic models, the tidal deformability  
fails to converge in this test unless the love number 
$k_2$ is computed using our improved implementation 
\Eref{eq:k2_robust_impl} instead of \Eref{eq:tidal_k2}.
The reason is that the compactness of those models decreases 
rapidly with polytropic index, leading to models more similar
to white dwarfs than neutron stars, thus triggering the cancellation
errors present for low compactness in the naive implementation. 

For practical purposes, it is important that users can specify 
the desired accuracy instead of technical details such as step size. 
For this, our implementation uses power laws for the residuals of 
each NS property as function of step size.  Those power laws
are chosen based on our test results, and are shown in 
\Fref{fig:accuracy_fixstep}. The exponents
are $1.6$ for the deformability and $1.8$ for all other quantities.
Those bounds are intended as heuristic estimates of the required 
resolution. We emphasize that one still needs to perform
a convergence test for a given model if reliable error bounds are 
required.

Our results show that the accuracy for fixed step size can vary
by orders of magnitude between different EOS. This suggests that
an adaptive step size might be more efficient. More importantly,
adaptive step size methods might reach the prescribed accuracy
also for corner cases not covered in our selection of test cases.
However, adaptive methods have one shortcoming for our use-case.
We recall that we first solve the TOV equations, while the
ODE for the tidal deformability is solved in a subsequent step.
Using an adaptive step size for the TOV equations will only reduce 
the step size as needed for those equations. The tidal ODE might 
require finer resolution of the TOV solution in very different 
locations, in particular at densities where the sound speed has 
sharp features.

For our main general-purpose implementation, we use a hybrid
approach for the step size selection.
Both for TOV and tidal ODE, we employ a RK4/5 ODE integrator with 
adaptive step size control. 
However, if the computation of the deformability is requested, 
we also enforce a minimum resolution for the integration of the 
TOV ODE. This resolution is based on the heuristic power law error 
model found for the fixed step size tests above.
Since ODE integration errors become less predictable in the 
low-accuracy regime, our implementation also employs a 
lower resolution limit.

\begin{figure*}
\caption{Accuracy of solution versus prescribed tolerance when using the 
default method.
The panels show the errors of different quantities: gravitational mass $M_g$,
baryonic mass $M_\mathrm{B}$, proper volume $V$, circumferential radius $R$,
moment of inertia $I$, and tidal deformability $\Lambda$. 
The tolerances not pertinent to the quantities shown in a given panel 
have no impact because they are set to a large value.
Note $R$ and $\sqrt[3]{V}$ share the same tolerance prescription, 
as do $M_g$ and $M_\mathrm{B}$.
The examples show results for many spline-based EOS, as well as a family
of polytropic EOS, and several piecewise polytropic ones.
Our tests also specified a minimum resolution limit of 20 points,
which causes the plateaus visible at low accuracy. 
}\label{fig:accuracy_adaptive}
\includegraphics[width=\textwidth]{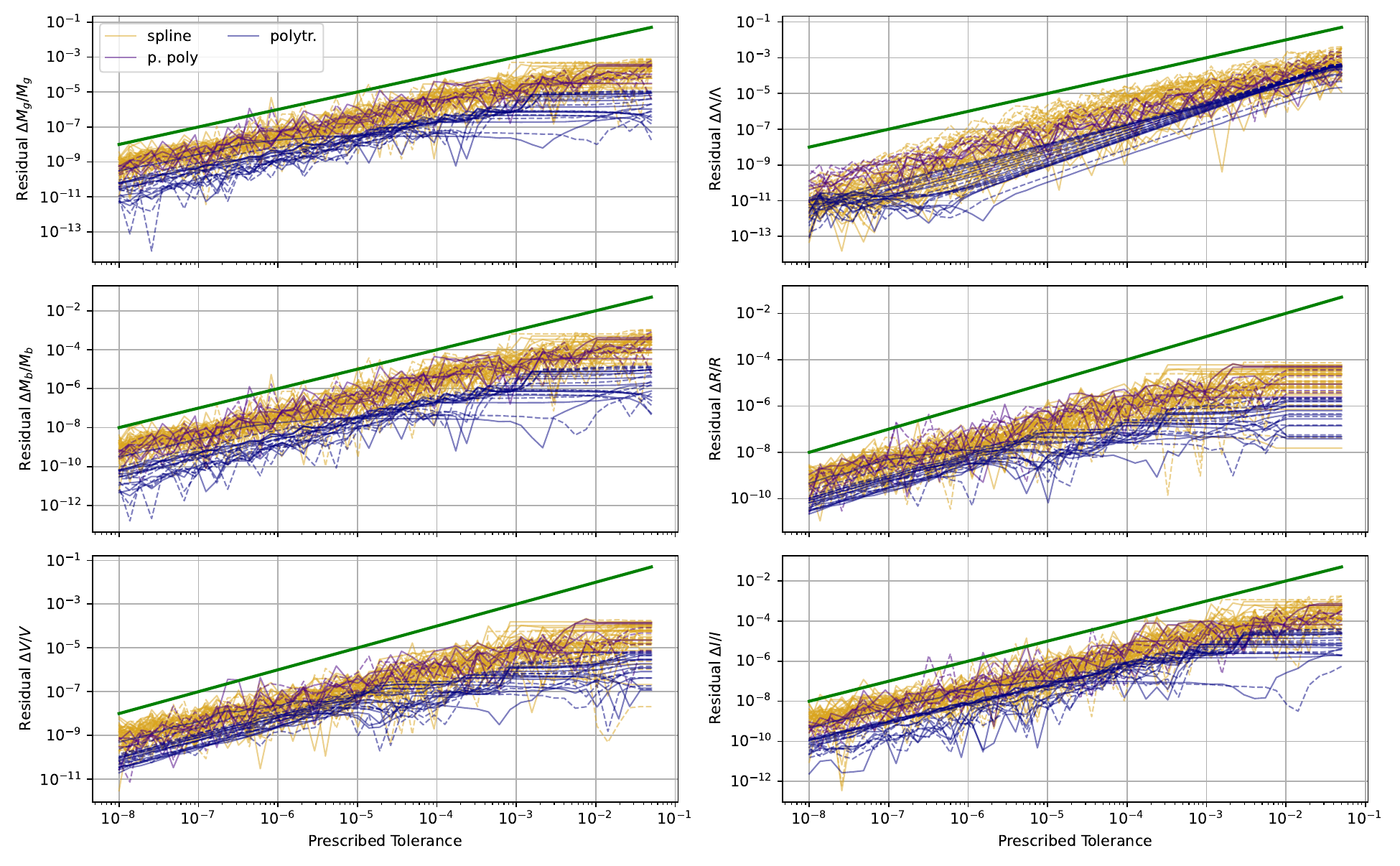}
\end{figure*}

In order to measure the accuracy and to 
calibrate the adaptive step size control, we solve the same 
models as for the previous test, but using different values 
for the local tolerance used in the step size control.
We then use simple constant scale factors to chose the 
local tolerance based on the desired global errors for the 
different NS properties. This allows to specify the desired
accuracy for each quantity separately.
\Fref{fig:accuracy_adaptive} shows the prescribed accuracy 
for each property in comparison to the measured residual
(again with respect to a much higher resolution).
As one can see, almost all models achieve the prescribed 
accuracy. As a trade-off between reliability and efficiency, 
we deliberately did not chose the calibration factors large 
enough to cover the few outliers. The error model is intended
as a heuristic guideline valid for typical EOS covered by 
our test cases. If more exact error bars are 
required, one should always perform a dedicated convergence 
test for the model at hand.

\subsection{Testing the Impact of EOS Approximation}
\label{sec:tests_eos}

Our next test concerns the inaccuracies introduced by approximating a
given EOS using monotonic spline interpolation. For this, we interpolate
several analytic EOS with increasing number of sample points and compute
the residuals of NS properties with respect to the original EOS.
The EOS used for this test are two piecewise polytropic EOS and one 
simple polytropic EOS. 

\begin{figure}
\caption{Errors of NS properties introduced by sampling a given EOS using
monotonic splines. The curves show the residual with respect to the original
EOS. Additionally, we show a measure specific to the ambiguity caused by the fact that
the sampling error is not the same when evaluating the EOS as function of density
or pseudo-enthalpy (see main text for details). 
The horizontal lines denote estimates for the impact of the low-density
approximation via polytropic EOS.
The figure contains results
for several piecewise polytropic EOS and a polytropic 
EOS with polytropic index $n=1$. 
}\label{fig:ambiguity_spline_eos}
\includegraphics[width=\columnwidth]{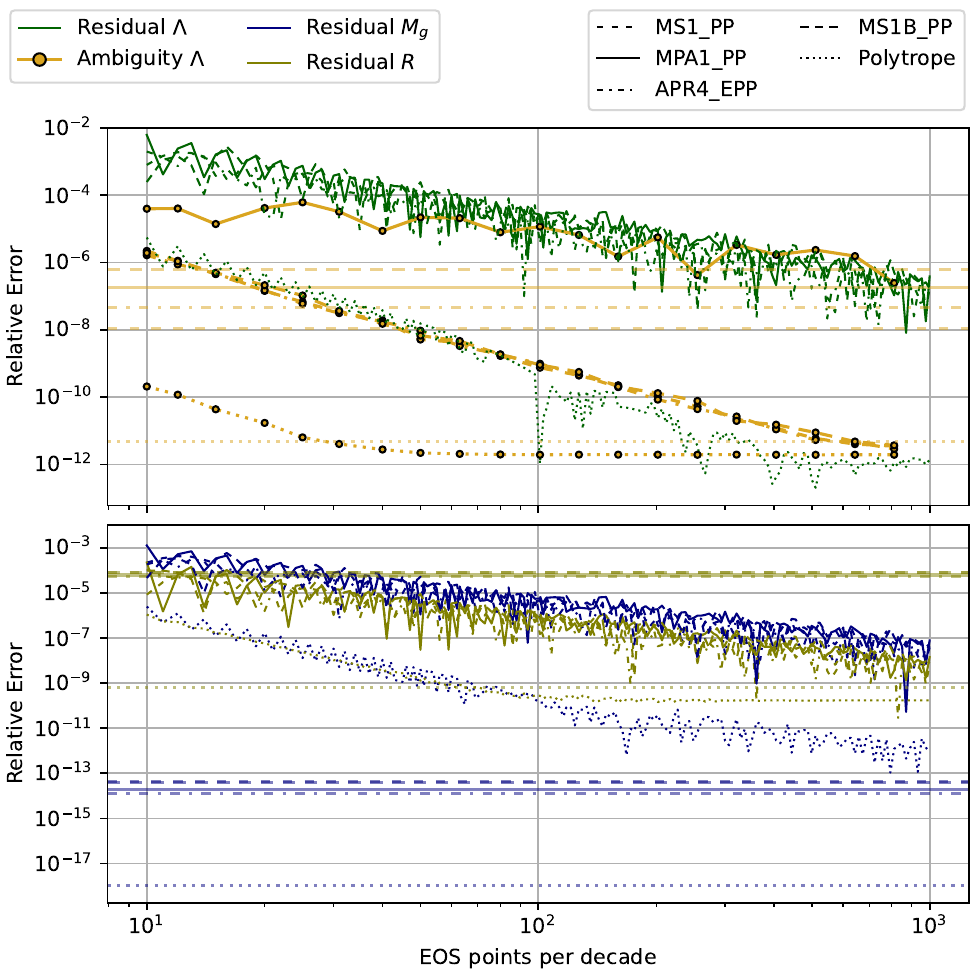}
\end{figure}

The residuals for mass, radius, and deformability are shown in \Fref{fig:ambiguity_spline_eos}. We find that the polytropic EOS 
is approximated much more accurate by the interpolation spline,
and converges faster. 
This is not surprising since the piecewise polytropic EOS is not 
differentiable at the segment boundaries, while the polytropic one is 
differentiable everywhere. 

However, for the polytropic case we also find that the residual of 
the radius does not decrease with resolution below a value of 
$10^{-9}$. As it turns out, the reason for this behavior is that
our implementation of the interpolating EOS uses a matching 
polytropic EOS below a given low density threshold $\rho_m$, in our example
at $\rho_m = \SI{1e8}{\kilo\gram\per\meter\cubed}$. The polytropic index of this
polytrope is chosen as $n \sim 1.7116$ for this test. This is the same value as 
the lowest segment of the piecewise polytropic examples, whereas
the purely polytropic EOS has a different index of $n=1$. 
At low densities, the polytropic example EOS is misrepresented, 
while the other two EOS are represented exactly.

We can easily obtain an estimate for the error introduced by the 
low-density approximation. Near the surface, we can assume
$P<E \ll m  / r^3$. Equations \Eref{eq:tov_dxdmu}, 
\Eref{eq:tov_m_of_lambda}, and \Eref{eq:tov_dhdr} then yield
the approximation
\begin{align}
\frac{\mathrm{d} \ln(x)}{\mathrm{d} \ln(H)}
&\approx -2 \frac{1-2\beta}{\beta}
\end{align}
where $\beta=M/R$. 
This allows to compute the thickness $\Delta R$ of the shell
with density below $\rho_m$. We find
\begin{align}\label{eq:NS_surf_thick}
\frac{\Delta R}{R}
&\approx \ln(H_m) \frac{1-2\beta}{\beta}
\end{align}
When using the polytropic approximation below $\rho_m$, 
and assuming that $H_m \equiv H(\rho_m) \ll 1$, we obtain
\begin{align}\label{eq:NS_radius_sys_err}
\frac{\Delta R}{R}
&\approx (n+1) \left(\frac{\rho_m}{\rho_p}\right)^{\frac{1}{n}} \frac{1-2\beta}{\beta}
\end{align}
The error in the NS radius caused by approximating the low density 
regime is given by the difference of $\Delta R$ obtained from 
\Eref{eq:NS_surf_thick} for original and approximating EOS. 
Computing $\Delta R$ from \Eref{eq:NS_radius_sys_err} already provides 
a useful estimate for the magnitude of the potential error, even though 
it is not a strict upper bound.
The results from the above error estimates are shown in 
\Fref{fig:ambiguity_spline_eos} as well. We find that the estimate
for the polytropic case roughly agrees with the observed
limitation of the NS radius accuracy.

It is worth pointing out that the error estimate for the other
examples is many orders of magnitude larger, even though
the matching density was the same in all cases. 
The actual error for this example is small compared to the estimate, 
but only because the low-density polytropic approximation happens 
to agree exactly with the original EOS. In general, however, this 
source of error needs to be taken into account.

The reason for the larger potential error is that the low-density
behavior of the polytropic example is very different from the other
examples. The polytropic constants $\rho_p$ was chosen such that the 
maximum NS mass is $2.2\, M_\odot$. As it turns out, the resulting
$\rho_p$ differs by orders of magnitude from the polytropic constant
of the lowest segment for the two piecewise polytropic EOS.
The latter are representative for realistic nuclear matter EOS,
for which the low-density behavior is relatively well constrained.

When processing many EOS from different sources, it is desirable to
automate the conversion into a spline representation. For this,
one may decide to always use the same low-density polytropic 
approximation mimicking realistic nuclear physics EOS, and to use 
the same matching density $\rho_m$. We can find an appropriate 
universal matching density using 
\Eref{eq:NS_radius_sys_err}
\begin{align}\label{eq:EOS_good_rho_m}
\rho_m &=
\rho_p 
\left(\frac{\beta_\text{min}}{n+1} \frac{\Delta R}{R}\right)^n
\end{align}
where $\Delta R / R$ is the desired accuracy of the radius,
and $\beta_\text{min}$ is the lowest compactness that will 
be considered.
When using a low density polytrope with $n=1.7115961$ and 
$\rho_p = \SI{9.53510e16}{\kilo\gram\per\meter\cubed}$,
we obtain a matching density 
$\rho_m = \SI{2e7}{\kilo\gram\per\meter\cubed}$
appropriate for $\beta_\text{min}= 0.06$ and $\Delta R / R = 10^{-4}$.

In contrast to the radius, the mass is not significantly affected
by the low-density behavior of the EOS. Approximating the 
TOV equations near the surface, we obtain an estimate
for the mass in the low density shell as
\begin{align}
\frac{\Delta M}{M} &\approx 
  4 \pi R^2 \frac{1-2 \beta}{\beta^2} P(\rho_m)
\end{align}
The result for our examples is shown in 
\Fref{fig:ambiguity_spline_eos}. As one can see, the impact 
on the mass is negligible.

Finally, we note that the use of interpolation splines to 
approximate EOS does not just introduce an error, but an ambiguity.
The reason is that $\rho(H)$ is not exactly the inverse function 
of $H(\rho)$, and thus one obtains slightly different results 
depending on which is used. One such ambiguity is manifest in
our implementation of the deformability. We recall that
we use two different formulations, one based on $\rho$ and one 
based on $H$, switching between the two at some point inside the 
star. To measure the above ambiguity, we use the standard deviation
of $\Lambda$ computed for 25 different choices (regularly spaced 
in $\log(H)$) for the location of the transition point.
\Fref{fig:ambiguity_spline_eos} shows the ambiguity 
versus the EOS sampling resolution. Again, we find much lower 
values for the polytrope. We also find that the ambiguity 
varies strongly within the piecewise polytropic examples,
for unknown reasons.

\section{Application}
\label{sec:application}

\subsection{Example EOS collection}
\label{sec:eoscoll}
As part of our EOS handling framework, we provide a small collection
of EOS files. The aim of this collection is twofold. First, we use it 
for our tests of the library. Second, it provides a convenient 
starting point for exploring various NS related questions for a
variety of EOS. The chosen set constitutes a representative selection
of nuclear physics EOS models. For a more extensive collection of 
EOS data, we refer to \cite{ComposeEOS}.
The full list of available EOS is given in \Tref{tab:eos_ns_props1} 
together with the maximum mass TOV solutions (see next section).
The EOS files are available in \cite{RePrimAnd:v1.7:EOS}.

In detail, our selection includes the EOS considered in 
\cite{LVC:EOSModelSel:2020}. Those EOS are based on tables available
in the literature on nuclear physics EOS modeling, but have been
sanitized by removing
clearly faulty samples, limiting the validity range to respect causality,
resampling coarsely sampled tables, and supplementing missing low-density 
data. Some EOS have also been replaced by analytic piecewise polytropic 
approximations from \cite{Read:2009:124032} because the original tables were 
sampled too coarsely for unambiguous interpolation. 
For details, we refer to \cite{LVC:EOSModelSel:2020}.

Our EOS framework
supports piecewise polytropic EOS natively. However, we also sampled 
some piecewise polytropic EOS to obtain representations based on spline 
interpolation. This was done only for testing purposes and the sampled 
variants are not used in the following sections.

The EOS used here are as close to the ones from \cite{LVC:EOSModelSel:2020}
as possible, but have been resampled to the regularly spaced values
employed by our EOS implementation. Note that the EOS data used in 
\cite{LVC:EOSModelSel:2020} was already resampled from the original 
sources to a suitable common resolution. Our example set is therefore not 
the closest possible representation of the original data, in particular 
with regard to sharp features, such as weak phase transitions.

\subsection{Properties of NS for Common EOS Models}
\label{sec:nsfits}

\begin{figure*}
\caption{NS properties along sequences of TOV solutions for the
EOS models listed in \Tref{tab:eos_ns_props1}, as function
of the gravitational mass. The curves for the different EOS are 
colored and sorted in the legend by the maximum NS mass.
Upper left: Circumferential radius. The dashed blue line divides
models that possess an ISCO (to the right) from those without (left).
Also shown is the Schwarzschild radius and radius of a BH photonsphere.
Upper right: Central baryonic mass density.
Middle left: Moment of inertia.
Middle right: Central speed of sound (note that the central value
does not always coincide with the maximum).
Bottom left: Tidal deformability.
Bottom right: Angular velocity of circular orbits at the NS surface.
Such orbits are unstable in the region shaded red, which is
bounded by the angular velocity at the ISCO for a BH.}
\label{fig:mass_radius}
\includegraphics[width=\textwidth]{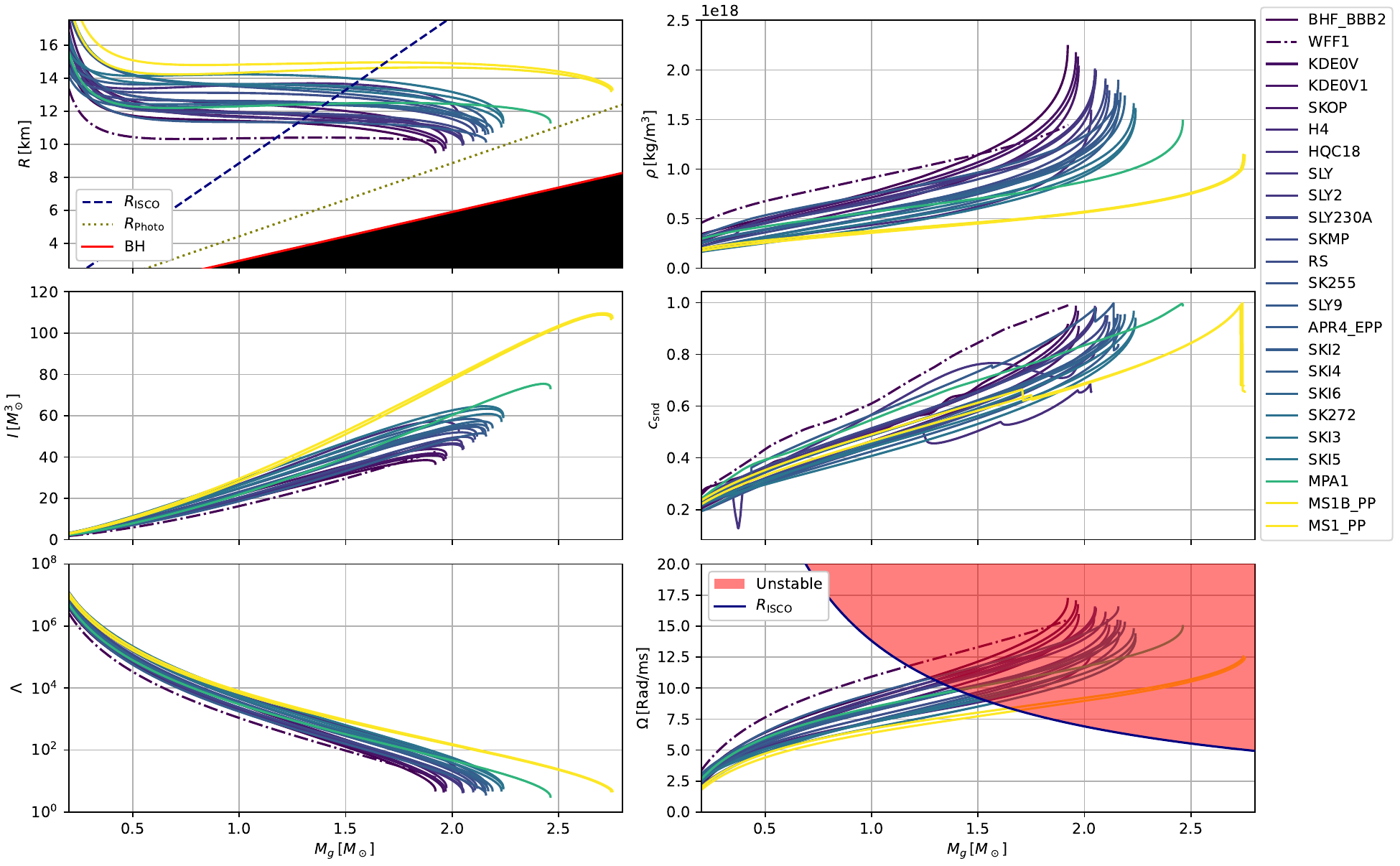}
\end{figure*}

As a first application of our code infrastructure, 
we compute the sequences of TOV solutions for our collection of EOS. 
\Fref{fig:mass_radius} shows the properties of stable NS as 
function of gravitational mass (up to the maximum). 

As one can see from the mass-radius relations, 
none of the sequences have a photonsphere, i.e. allow 
circular photon orbits outside the star.
Further, all of the TOV sequences in our selection develop an 
ISCO before reaching the maximum mass. 
This can also be seen from the panel showing the angular
velocity of circular orbits at the NS surface in comparison to 
the one at the ISCO radius. In general, the surface orbital 
angular velocity increases with mass for our selection.
Once it crosses the ISCO angular velocity, which decreases with mass, 
the surface orbits are unstable.

The panel showing the central density is useful to interpret 
EOS constraints from GW observations, since the signal cannot be 
influenced by the EOS at densities above the central ones for
given range of involved NS masses. Of course, it is still possible
to infer constraints on the pressure beyond those densities,
using the causality constraint on the speed of sound.
The mass-density plot also shows that the maximum NS mass and 
the central density of the corresponding NS are anticorrelated 
for our selection.

The mass-central soundspeed relation plot exhibits non-monotonic
behavior as well as sharp features.
The reason is simply that EOS can have a non-monotonic 
soundspeed-density relation and sharp features. For the same 
reason, the maximum sound speed inside a NS is not always the 
central value.

\Fref{fig:mass_radius} also shows the moment of inertia.
For any search for GW from single rotating NS, the moment 
of inertia $I$ is an important quantity as it relates GW amplitude,
GW frequency, and spindown rate (for GW-dominated spindown). 
The plot can be used to assess the error made when using ballpark 
figures for typical NS. It also shows that maximum mass and maximum moment
of inertia are strongly correlated.

The mass-deformability plot shows that the maximum mass models
for our selection generally have a low tidal deformability of order 10.
For current ground-based detectors, those maximum mass NS models are therefore
near undistinguishable from BH. This seems to be mainly related to the high 
compactness of the maximum mass models for our EOS selection. 
As we will see in \Sref{sec:splitting}, one can easily construct EOS that have
mass maxima with lower compactness and much higher tidal deformability.

For each EOS, the NS model of maximum mass is of particular interest.
\Tref{tab:eos_ns_props1} provides
gravitational mass $M_g$, baryonic mass $M_b$, proper circumferential 
radius $R_c$, moment of inertia $I$, central baryonic mass 
density $\rho_c$, and central sound speed $c_s$.
In addition, we compute novel measures introduced
in \cite{Kastaun:2016}, dubbed ``bulk mass'' and ``bulk compactness''
(see \Sref{sec:bulkmeasures}).
The bulk measures of the 
maximum mass TOV solution may be useful because they appear
in a recently proposed empirical criterion for post-merger 
BH formation \cite{Ciolfi:2017:063016,Endrizzi:2018,Kastaun21:023001}.
We checked that our results on the maximum masses are 
consistent with the ones reported
in \cite{LVC:EOSModelSel:2020}.

Next, we compute NS 
properties at a fiducial mass $M_g^\mathrm{fid}=1.4 M_\odot$. 
\Tref{tab:eos_ns_props2} provides 
baryonic mass $M_b$, proper circumferential 
radius $R_c$, central baryonic mass density $\rho_c$, moment of 
inertia $I$, the angular velocity of internal circular orbits 
near the center according to \Eref{eq:omega_orb_center}, 
and dimensionless tidal deformability $\Lambda$.
In addition, we provide the first derivative 
\begin{align} 
  S(M_g) \equiv \frac{\mathrm{d} \ln(\Lambda)}{\mathrm{d}\ln(M_g)}
  \left(M_g\right)
\end{align}
evaluated at $M_g = M_g^\mathrm{fid}$.

Expanding the logarithmic  tidal deformability $\ln(\Lambda(M_g))$ 
around the fiducial mass to first order might be sufficient for
many applications. For example, \cite{SoumiDe:2018} analyzed the 
GW data for the single event GW170817  
assuming a constant slope of $S=-6$ for all EOS and kept only the 
deformability at some fiducial mass as a free parameter.
This approximation thus reduces the
infinite-dimensional space of EOS to a single degree of freedom.
Once there are further constraints on $\Lambda$ at 
different masses from future observations, it will become 
possible to constrain two 
degrees of freedom, which can be expressed in terms of  
$\Lambda$ and $S$ at some fixed fiducial mass.
We caution that Taylor expansion (to any order)
ceases to be a viable approach when considering EOS
that lead to multiple stable NS branches. This case 
will be discussed in \Sref{sec:splitting}.

\begin{table*}
\caption{Properties of the maximum mass nonrotating NS model for various EOS. 
For each EOS, we provide gravitational mass $M_g$, baryonic mass $M_b$, proper circumferential radius $R_c$, the ``bulk mass'' $M_b^\text{blk}$ and
``bulk compactness'' $C^\text{blk}$ defined in \cite{Kastaun:2016}, 
moment of inertia $I$, central baryonic mass 
density $\rho_c$, and central sound speed $c_s^\text{cnt}$.
EOS marked by a star become physically 
invalid (violating causality by superluminal sound speed) at a 
density exceeded within NSs before reaching the mass maximum.
The values then refer to the model with the maximal
central density that is still physically valid.}
\begin{tabular}{lSSSSSSSS}
{EOS} & {$M_g \,[M_\odot]$} & {$M_b\,[M_\odot]$} & {$R_c \,[\si{\kilo\meter}]$} & {$M_b^\text{blk}\,[M_\odot]$} & {$C^\text{blk}$} & {$I \,[M_\odot^3]$} & {$\rho_c\,[\SI{1e18}{\kilo\gram\per\meter\cubed}]$} & {$c_\mathrm{csnd}^\mathrm{cnt}\,[c]$} \\
\hline
BHF\_BBB2 \cite{1997AA...328..274B} & 1.9214 & 2.2683 & 9.5222 & 2.2132 & 0.3212 & 36.764 & 2.2406 & 0.9139 \\
*WFF1 \cite{1988PhRvC..38.1010W} & 1.9235 & 2.3059 & 10.196 & 2.2671 & 0.3131 & 42.523 & 1.4458 & 0.9902 \\
KDE0V \cite{2015PhRvC..92e5803G,2005PhRvC..72a4310A,2009NuPhA.818...36D} & 1.9600 & 2.3130 & 9.6577 & 2.2473 & 0.3229 & 38.658 & 2.1671 & 0.9843 \\
KDE0V1 \cite{2015PhRvC..92e5803G,2005PhRvC..72a4310A,2009NuPhA.818...36D} & 1.9693 & 2.3175 & 9.7924 & 2.2474 & 0.3199 & 39.526 & 2.1241 & 0.9648 \\
SKOP \cite{2015PhRvC..92e5803G,1999PhRvC..60a4316R,2009NuPhA.818...36D} & 1.9727 & 2.3042 & 10.125 & 2.2185 & 0.3089 & 41.062 & 2.0325 & 0.9059 \\
H4 \cite{2006PhRvD..73b4021L} & 2.0314 & 2.3413 & 11.735 & 2.2637 & 0.2751 & 52.518 & 1.5886 & 0.6545 \\
HQC18 \cite{2018RPPh...81e6902B} & 2.0450 & 2.4265 & 10.387 & 2.3488 & 0.3181 & 46.473 & 1.8823 & 0.8035 \\
SLY \cite{2001AA...380..151D} & 2.0490 & 2.4286 & 9.9927 & 2.3660 & 0.3265 & 43.919 & 2.0030 & 0.9836 \\
SLY2 \cite{2015PhRvC..92e5803G,2009NuPhA.818...36D} & 2.0535 & 2.4334 & 10.045 & 2.3694 & 0.3256 & 44.352 & 1.9874 & 0.9770 \\
SLY230A \cite{2015PhRvC..92e5803G,1997NuPhA.627..710C,2009NuPhA.818...36D} & 2.0988 & 2.4966 & 10.251 & 2.4412 & 0.3272 & 47.571 & 1.9013 & 0.9477 \\
SKMP \cite{2015PhRvC..92e5803G,1989PhRvC..40.2834B,2009NuPhA.818...36D} & 2.1069 & 2.4829 & 10.527 & 2.3997 & 0.3171 & 48.947 & 1.8378 & 0.9465 \\
RS \cite{2015PhRvC..92e5803G,1986PhRvC..33..335F,2009NuPhA.818...36D} & 2.1164 & 2.4807 & 10.763 & 2.3855 & 0.3108 & 50.404 & 1.7832 & 0.9223 \\
SK255 \cite{2015PhRvC..92e5803G,2003PhRvC..68c1304A,2009NuPhA.818...36D} & 2.1439 & 2.5097 & 10.849 & 2.4150 & 0.3132 & 51.847 & 1.7541 & 0.9371 \\
SLY9 \cite{2015PhRvC..92e5803G,2009NuPhA.818...36D} & 2.1558 & 2.5518 & 10.634 & 2.4842 & 0.3228 & 51.858 & 1.7825 & 0.9523 \\
APR4\_EPP \cite{2009PhRvD..79l4032R,PhysRevC.58.1804,Endrizzi:2016:164001} & 2.1589 & 2.6105 & 10.171 & 2.5575 & 0.3433 & 50.467 & 1.8897 & 0.8360 \\
SKI2 \cite{2015PhRvC..92e5803G,1995NuPhA.584..467R,2009NuPhA.818...36D} & 2.1627 & 2.5262 & 11.114 & 2.4245 & 0.3070 & 54.284 & 1.6857 & 0.9139 \\
SKI4 \cite{2015PhRvC..92e5803G,1995NuPhA.584..467R,2009NuPhA.818...36D} & 2.1693 & 2.5770 & 10.670 & 2.5126 & 0.3235 & 52.971 & 1.7619 & 0.9493 \\
SKI6 \cite{2015PhRvC..92e5803G,1996PhRvC..53..740N,2009NuPhA.818...36D} & 2.1897 & 2.6011 & 10.762 & 2.5365 & 0.3239 & 54.408 & 1.7315 & 0.9513 \\
SK272 \cite{2015PhRvC..92e5803G,2003PhRvC..68c1304A,2009NuPhA.818...36D} & 2.2314 & 2.6275 & 11.086 & 2.5393 & 0.3195 & 57.532 & 1.6548 & 0.9645 \\
SKI3 \cite{2015PhRvC..92e5803G,1995NuPhA.584..467R,2009NuPhA.818...36D} & 2.2397 & 2.6307 & 11.309 & 2.5424 & 0.3134 & 59.348 & 1.6056 & 0.9373 \\
SKI5 \cite{2015PhRvC..92e5803G,1995NuPhA.584..467R,2009NuPhA.818...36D} & 2.2399 & 2.6152 & 11.467 & 2.5041 & 0.3077 & 59.901 & 1.5822 & 0.9317 \\
MPA1 \cite{1987PhLB..199..469M} & 2.4619 & 3.0117 & 11.325 & 2.9683 & 0.3527 & 73.305 & 1.4880 & 0.9899 \\
MS1B\_PP \cite{1996NuPhA.606..508M,2009PhRvD..79l4032R} & 2.7463 & 3.3080 & 13.224 & 3.2569 & 0.3349 & 106.79 & 1.1431 & 0.6795 \\
MS1\_PP \cite{1996NuPhA.606..508M,2009PhRvD..79l4032R} & 2.7528 & 3.3029 & 13.312 & 3.2386 & 0.3324 & 107.47 & 1.1375 & 0.6555 \\
\hline
\end{tabular}
 \label{tab:eos_ns_props1}
\end{table*}

\begin{table*}
\caption{Properties of NS models with a fiducial gravitational mass 
$M_g = 1.4\,M_\odot$, for various EOS.
For each EOS, we provide baryonic mass $M_b$, proper circumferential 
radius $R_c$, central baryonic mass density $\rho_c$, moment of 
inertia $I$, angular velocity of internal circular orbits near the 
center, dimensionless tidal deformability $\Lambda$, and the 
derivative $S\equiv \mathrm{d} \ln(\Lambda)/\mathrm{d}\ln(M_g)$ 
evaluated at the above fiducial mass.}
\begin{tabular}{lSSSSSSS}
{EOS} & {$M_b\,[M_\odot]$} & {$R\,[\si{\kilo\meter}]$} & {$\rho_c\,[\SI{1e18}{\kilo\gram\per\meter\cubed}]$} & {$I \,[M_\odot^3]$} & {$\Omega_c^K \,[\si{\radian\per\milli\second}]$} & {$\Lambda / 100$} & {$S$} \\
\hline
BHF\_BBB2 & 1.5545 & 11.175 & 1.0519 & 29.085 & 13.51 & 2.1598 & -7.324 \\
WFF1 & 1.5787 & 10.407 & 1.0997 & 26.731 & 13.64 & 1.5103 & -6.451 \\
KDE0V & 1.5501 & 11.425 & 0.98955 & 29.901 & 13.20 & 2.4161 & -7.056 \\
KDE0V1 & 1.5461 & 11.634 & 0.95501 & 30.627 & 13.03 & 2.6626 & -7.045 \\
SKOP & 1.5401 & 12.137 & 0.86340 & 33.074 & 12.53 & 3.6127 & -7.108 \\
H4 & 1.5313 & 13.686 & 0.57159 & 42.239 & 10.53 & 8.9915 & -6.182 \\
HQC18 & 1.5504 & 11.492 & 0.90063 & 30.367 & 12.70 & 2.5684 & -6.369 \\
SLY & 1.5461 & 11.724 & 0.89384 & 31.499 & 12.68 & 2.9719 & -6.719 \\
SLY2 & 1.5460 & 11.792 & 0.88055 & 31.829 & 12.61 & 3.0969 & -6.707 \\
SLY230A & 1.5467 & 11.841 & 0.84314 & 32.345 & 12.38 & 3.2942 & -6.476 \\
SKMP & 1.5382 & 12.507 & 0.75220 & 35.588 & 11.84 & 4.7761 & -6.677 \\
RS & 1.5329 & 12.942 & 0.69853 & 37.648 & 11.50 & 5.9064 & -6.732 \\
SK255 & 1.5264 & 13.157 & 0.69490 & 37.547 & 11.49 & 5.8518 & -6.697 \\
SLY9 & 1.5370 & 12.478 & 0.74822 & 35.034 & 11.81 & 4.4902 & -6.419 \\
APR4\_EPP & 1.5542 & 11.320 & 0.91778 & 30.112 & 12.78 & 2.4761 & -6.359 \\
SKI2 & 1.5265 & 13.496 & 0.62532 & 40.446 & 10.99 & 7.6885 & -6.624 \\
SKI4 & 1.5419 & 12.383 & 0.73372 & 35.434 & 11.70 & 4.6848 & -6.342 \\
SKI6 & 1.5398 & 12.500 & 0.71763 & 35.856 & 11.60 & 4.8993 & -6.298 \\
SK272 & 1.5250 & 13.326 & 0.65444 & 38.495 & 11.20 & 6.4090 & -6.443 \\
SKI3 & 1.5256 & 13.567 & 0.60209 & 40.678 & 10.81 & 7.8355 & -6.335 \\
SKI5 & 1.5205 & 14.096 & 0.55726 & 43.601 & 10.47 & 10.077 & -6.528 \\
MPA1 & 1.5450 & 12.455 & 0.67292 & 35.831 & 11.27 & 4.8760 & -5.766 \\
MS1B\_PP & 1.5161 & 14.528 & 0.44505 & 46.143 & 9.497 & 12.243 & -5.412 \\
MS1\_PP & 1.5112 & 14.926 & 0.43110 & 47.726 & 9.377 & 13.793 & -5.524 \\
\hline
\end{tabular}
 \label{tab:eos_ns_props2}
\end{table*}

\subsection{Multimessenger Applications}
\label{sec:multimess}

As another simple application of our code, we consider the case of
a multi-messenger BNS merger detection where EM counterparts point 
to the formation of a BH.
We will discuss some simple consequences of the assumption that a 
BH was formed, which would need to be considered
in Bayesian parameter estimation and model selection studies
of the GW data.

For a given EOS, we can easily compute the total baryonic
mass of the system from the gravitational masses of the constituents,
as parameterized by chirp mass $M_c$ and mass ratio $q=M_2/M_1$.
For simplicity, we assume that spin effects can be neglected.
It is also straightforward to compute the total baryonic mass
that can be contained in a single nonrotating NS.
It is a safe assumption that the remnant is a stable NS if the total
baryonic mass is below the above value.
We note that although \textit{a priori} the merger could still 
produce a BH by some dynamical effect, we are not aware of any 
such example from numerical relativity simulations.
It should also be noted that BNS merger remnants have considerable 
angular momentum and can therefore support masses larger than
the maximum for nonrotating NS, forming a long-lived supramassive 
remnant.

\begin{figure}
\caption{Maximum chirp mass at which the total baryonic mass within
a BNS system of two nonrotating NS is below the maximum baryonic mass
admissible for a single nonrotating NS, as function of mass ratio.
The relation is computed for many EOS, which are sorted in the legend  
by maximum NS mass. The chirp masses provide a reliable (but not very tight) 
lower limit for the formation of a BH any time after the merger.
The horizontal line marks the chirp-mass estimate inferred for GW170817
\cite{LVC:BNSSourceProp:2019}.}
\label{fig:stable_remnant}
\includegraphics[width=\columnwidth]{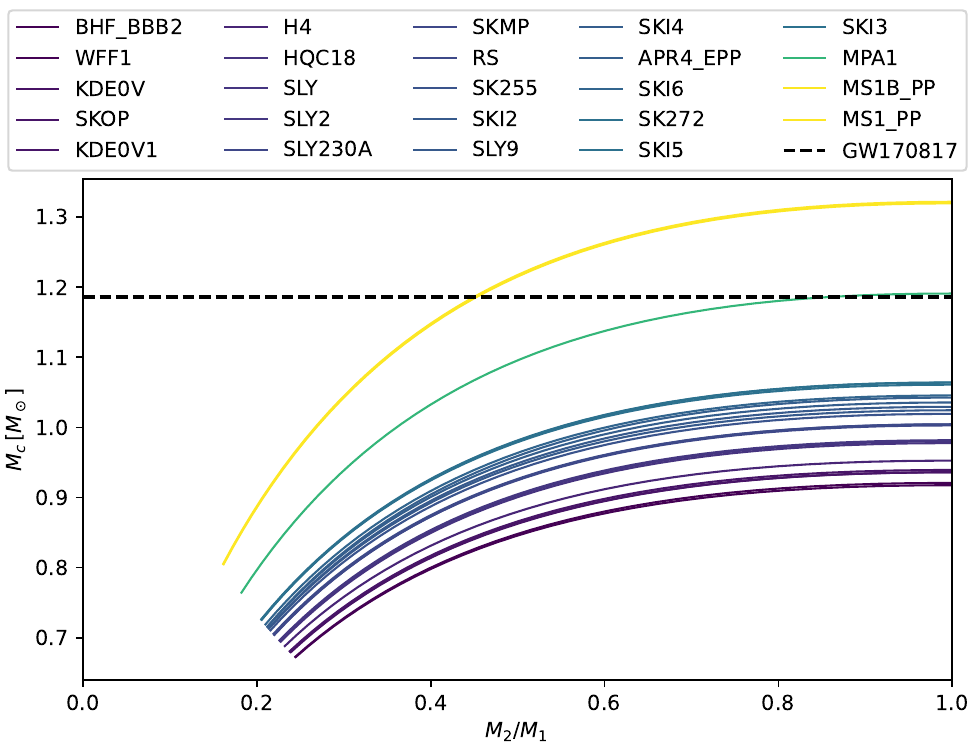}
\end{figure}

In \Fref{fig:stable_remnant}, we show the resulting lower 
chirp mass limit for BH formation as function of mass ratio. 
For comparison, we show the estimate for the chirp mass in 
GW event GW170817 \cite{LVC:BNSSourceProp:2019}. 
For this and similar events, the chirp mass can be constrained 
so well that for our discussion we can assume it to be known 
exactly. From the figure, we can read off immediately that
the knowledge of the chirp mass can place strong constraints
on the mass ratio, if the EOS is given. Under the assumption that a BH has formed
the possible mass ratio drops rapidly once the chirp mass
drops below a critical (EOS-dependent) value.
The BNS  mass ratio is an important nuisance parameter in GW 
parameter estimation studies aiming to constrain the EOS. 
Hence it might be beneficial to include the assumption of BH 
formation in the prior, by excluding at least the region below
the aforementioned chirp mass limit. Of course, the
correct approach is to use estimates for BH formation thresholds 
instead of simple lower limits, which could be obtained from numerical 
relativity simulations. 

Computing the excluded region in the mass prior requires knowledge
of the EOS. For Bayesian model selection studies, the EOS is given.
For parameter estimation studies employing parametrized EOS,
one needs to consider a different mass prior for each EOS.
In other words, the assumption of BH formation introduces 
an additional correlation between priors for masses and EOS parameters,
besides the existence of a maximum NS mass.

\subsection{Universal Relations for GW}
\label{sec:universal}

There are several observational avenues for constraining the EOS 
of NS matter. A detection of GW from a BNS coalescence mainly provides 
constraints in the $(M, \Lambda)$ plane (setting aside complications when 
allowing for rapid spins or unequal masses) but no direct information
on the radius (this might change once a signal from the merger itself 
can be observed). Electromagnetic observations such as NICER 
\cite{NICER:Bogdanov:2019:L25, NICER:2016SPIE} provide constraints in 
$(M,R)$, but no direct measurement of $\Lambda$.

Both $\Lambda$ and $R$ depend on the EOS, so it is ---in principle--- 
possible to obtain combined EOS constraints in a fully consistent 
Bayesian manner. The difficulty lies in the 
nontrivial parametrization of the EOS uncertainties and the requirement
for fully self-consistent prior assumptions. 

The situation would be much simpler if radius and deformabilty would depend 
on the EOS in the same way. In detail: for a given mass, each
possible EOS corresponds to a point in the $(R,\Lambda)$ plane. 
Let us assume that the set of those points does not constitute a 
two-dimensional region, but is constrained to some one-dimensional 
curve (or at least a sufficiently narrow band). Further, assume
that this curve can be described as a monotonic function $R(\Lambda)$.
One can then convert measurements of $\Lambda$ into measurements 
of $R$, and vice versa, without considering the EOS.

There are several proposals for such EOS-independent relations
(universal relations) \cite{Yagi:2017:1,SoumiDe:2018} 
given as a functional relation
$\beta(\Lambda)$, where $\beta=M_g/R$ is the compactness.
We note that this relation is more restrictive than required. 
For the purpose
of simplifying combined data analysis, it would be sufficient 
to have a relation $\beta_M(\Lambda)$ for each mass $M$,
whereas the above universal relation $\beta(\Lambda)$ is 
the same for all masses.

\begin{figure}
\caption{Relation between compactness and love number $k_2$ for
various EOS, together with the ``universal'' relations from 
\cite{Yagi:2017:1, SoumiDe:2018}.
In addition to a large selection of nuclear physics EOS models (see \Sref{sec:eoscoll})
we plot results for the unrealistic polytropic EOS used for testing 
(see \Sref{sec:tests_eos}).}
\label{fig:compact_lambda}
\includegraphics[width=\columnwidth]{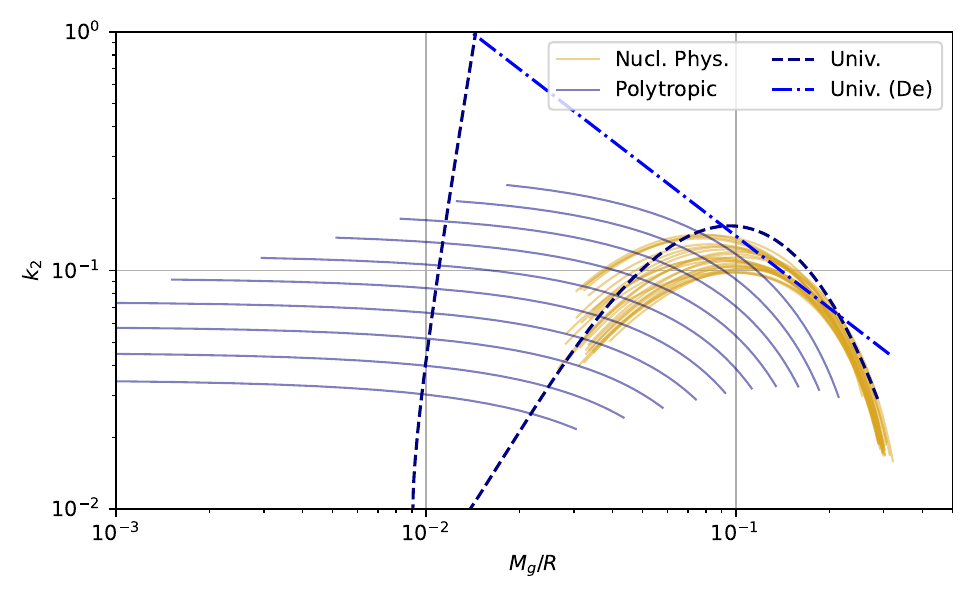}
\end{figure}

Both universal relations from \cite{Yagi:2017:1, SoumiDe:2018} 
are derived as fits to a small
selection of EOS models predicted by nuclear physics. 
Even for those EOS models considered, the relations have 
considerable residuals and can be called quasi-universal 
relations at best. 
This can be seen in \Fref{fig:compact_lambda} showing the relation
computed with our code for many nuclear physics EOS. 
Instead of $\beta(\Lambda)$, we show the equivalent relation 
$k_2(\beta)$. As one can see, neither universal relation 
fits the band spanned by the EOS particularly well.
The plot highlights that the ``universality'' of 
$\beta(\Lambda)$ is simply owed to the factor $\beta^5$ 
in \Eref{eq:tidal_def_lambda}, i.e., the strong dependency on 
the compactness dominates other factors.

Using the universal relation to constrain the EOS from given 
observational data implies the assumption that the relation is 
respected by all EOS not yet ruled out by previous data, 
and not just by the selected examples used to establish 
the relation. In other words, any constraints
on the EOS would already be based on assumptions about the EOS.
We note that universal relations have already been used to 
convert deformability constraints into radius constraints
in the context of gravitational wave event GW170817
\cite{LVC:EOSPaper:2018, SoumiDe:2018}.

Showing that the universal relations are truly universal in that sense
is difficult. However, it is easy to answer the question whether
the universal relations are a good approximation for all EOS, 
realistic or not. The set of polytropic EOS used for our tests
provides counterexamples, as can be seen in \Fref{fig:compact_lambda}.
We stress that those polytropic EOS differ from the nuclear physics 
models already at low density, where the EOS is well constrained.

To get a more systematic measure for the universality of universal relations, 
we used our library to perform a numerical search for larger violations 
of said relations over a family of parametrized EOS.
Our motivation is to get the universal relation violations under 
the simplistic assumption that the EOS at low densities
is constrained by the spread of available nuclear physics models, while 
at high densities the EOS is completely unconstrained.
The parametrized EOS we use to this end are based on modifying 
tabulated nuclear physics EOS only above a density threshold.
For densities above $\SI{5E16}{\kilo\gram\per\meter\cubed}$, 
we replace the original speed of sound by
a smooth monotonic interpolation between a small number of sample
points at fixed energy densities. The parameter space of the EOS
is simply given by the sound speeds at those points.
From the speed of sound, the full EOS
is constructed using \Eref{eq:eos_adiabatic} and 
\Eref{eq:eos_csnd_adiab} valid for isentropic (cold) EOS.

We then use a simple optimization scheme to look for the parameters 
with largest deviations from the universal relations. For this, 
we maximize a scalar function that computes the $L_2$-norm 
of the compactness deviation as function of mass,
weighted by a Gaussian mass ``prior''. 
Doing so, we found that this can lead to parameters for which 
the maximum NS mass
is well below $2\,M_\odot$. To suppress this region of parameter space,
we multiply the scalar function by a penalty factor that is
a smoothed-out step function in terms of the maximum NS mass.
Another complication which emerged is the splitting of the stable
NS branch into two, separated by an unstable branch. To account
for this, we use the larger of the two mass maxima in the penalty
factor for the maximum mass.
Further, for a mass within the overlap of the mass ranges spanned 
by two branches, there are also two different values for the 
violation of the universal relation. We simply use the mean of the 
squares in the $L_2$-norm of the scalar function we maximize.

We note that our maximization scheme is not designed to 
reliably find global maxima inside the multi-dimensional parameter 
space. Also note that we optimize the average violations and 
one could probably find larger violations for any given particular 
mass. Further, our EOS parametrization is not well suited to model
phase transitions. As shown in \cite{Han:2019:083014}, strong phase 
transitions can also have a significant effect on the deformability.
The results of our search should therefore be regarded as examples
that provide lower limits for the possible violations.

\begin{figure}
\caption{Tidal deformability versus compactness for example set
of EOS, for three selected masses. The filled markers denote results for nuclear 
physics EOS, while the plus markers denote results for the same EOS modified
in the high-density region. Those were found by a numeric search for large 
violations of the universal relations while allowing a NS mass of $2\,M_\odot$
(see main text). For comparison, we show universal relations from 
\cite{LVC:EOSPaper:2018} (labeled ``Univ.'') and \cite{SoumiDe:2018}
(``Univ (De)''.}
\label{fig:violate_univ}
\includegraphics[width=\columnwidth]{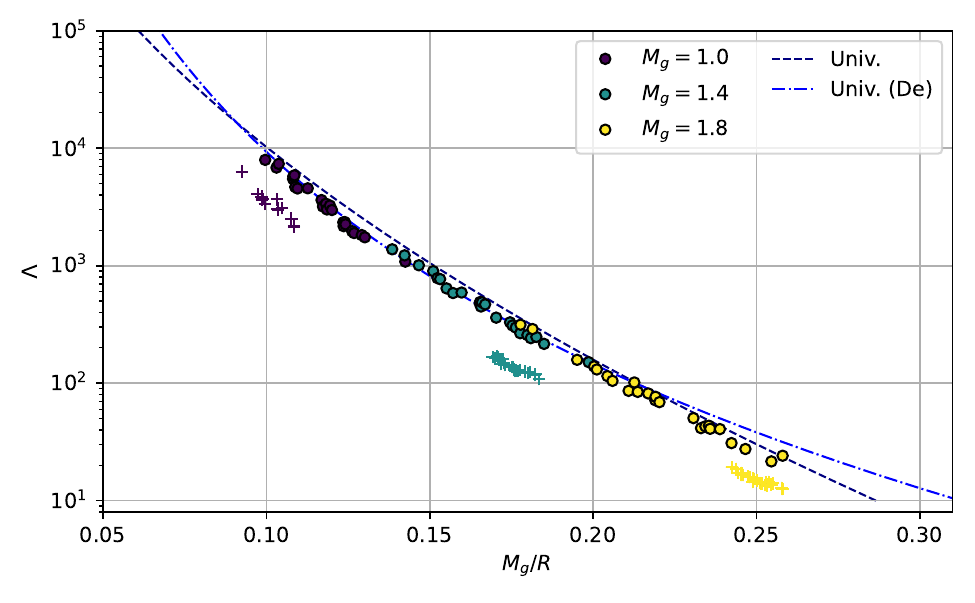}
\end{figure}

The results for parametrized EOS based on each of our nuclear physics EOS
examples are shown in \Fref{fig:violate_univ}.
For each EOS, it displays tidal deformabilty versus compactness for three 
selected masses, both for the original EOS and the corresponding 
parametrized EOS with the largest overall violation found by our search.
As one can see, the deviations from the universal relations
can be substantially larger than the spread of the original nuclear 
physics EOS. 
\emph{Our findings show that the compactness-deformability 
universal relations cannot be used in any study aiming to constrain 
the high-density part of the EOS}, since obviously one already 
needs to assume EOS constraints to satisfy those relations.

\begin{figure}
\caption{Neutron star TOV sequences for a 1-parameter family of 
parametrized EOS that gradually modifies the MPA1 EOS, keeping the 
low-density part fixed (see main text). 
If there is only one stable branch for a given member of the 
EOS family, it is shown with black solid 
curves, while red and green solid curves show multiple stable 
branches. The yellow dashed curve marks unstable branches,
and the solid blue curve shows the full TOV sequence for 
the original MPA1 EOS.
Top: Gravitational mass versus radius.
Middle: Tidal deformability versus gravitational mass.
Bottom: Soundspeed versus density for the original EOS and 
members of the parametrized EOS family. The vertical lines
mark the central densities for NS with the original EOS 
and masses $1.0, 1.4, 1.8 \,M_\odot$ as well as the maximum mass
NS.}
\label{fig:parametrized_eos}
\includegraphics[width=\columnwidth]{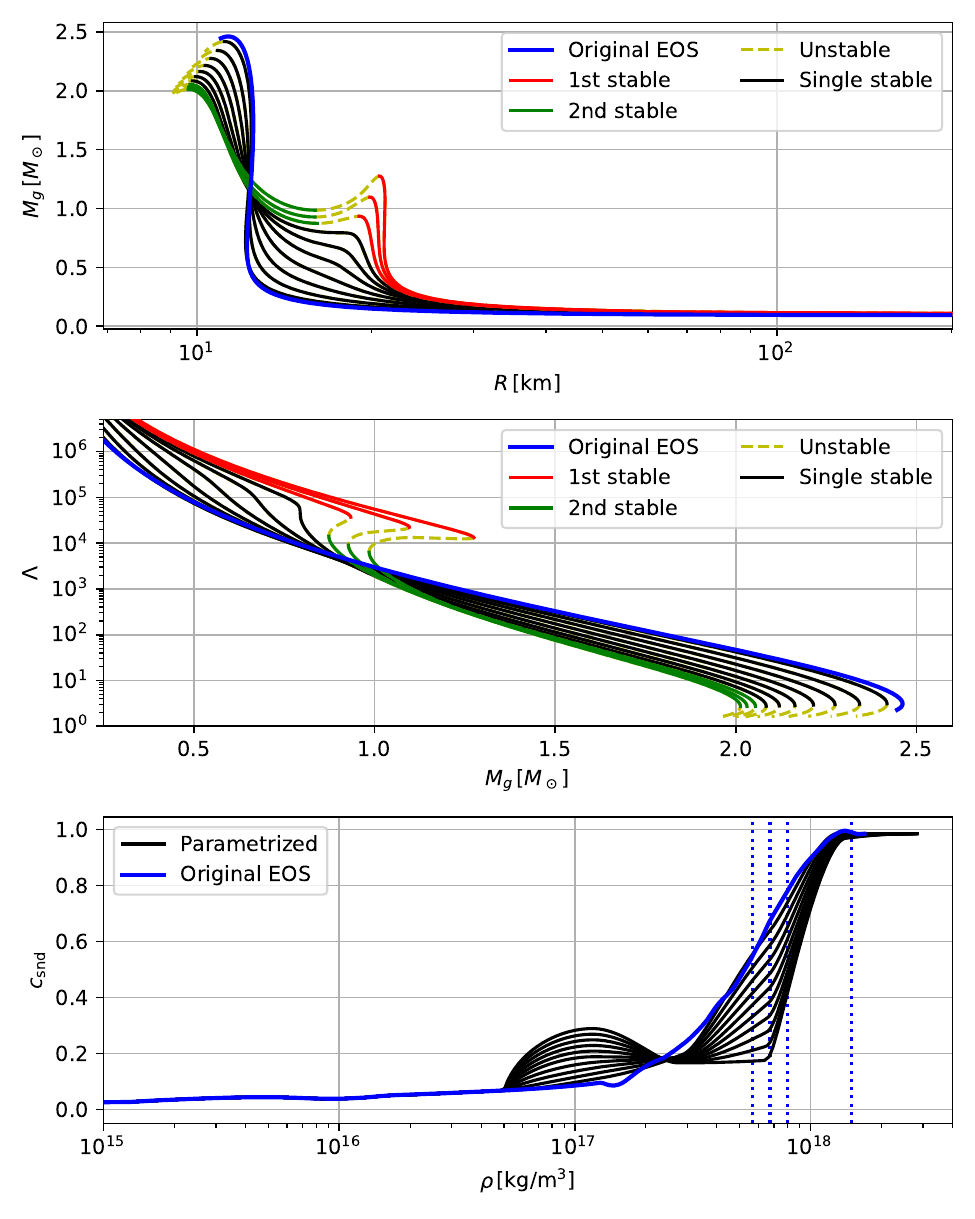}
\end{figure}

It is instructive to compare the parametrized EOS 
members that violate the universal relations to the original EOS.
This is shown in the bottom panel of \Fref{fig:parametrized_eos}.
Apparently, decreasing the soundspeed at high density while increasing
it at medium densities leads to the largest deviations found
in our search.
It should be noted that the spread surely depends on the 
density $\rho_\mathrm{fix}$ below which the EOS is kept constant.
Our choice is pretty low, around one quarter of the nuclear saturation 
density. Hence, \emph{our results are not valid counterexamples against 
the use of universal relations for applications where the EOS is 
already assumed to be constrained by other means up to higher 
densities.}

We remark that \Fref{fig:violate_univ} highlights another problem
with using the universal relations. As we already pointed out in 
\cite{Kastaun:2019}, the range of compactness at a given mass is limited,
at least when considering only the EOS used to calibrate the universal 
relations.
It is therefore not self-consistent to simply use universal relations 
of the form $\beta(\Lambda)$ to convert a posterior probability density 
for $\Lambda, M$ that was obtained from some generic prior into a 
posterior for $R,M$. 
For example, applying the universal relation from \cite{SoumiDe:2018}  
to a parameter sample with $M=M_\odot$ and $\Lambda=0.01$
is not meaningful because this region of parameter 
space is clearly excluded for all EOS models used to derive the universal 
relation (moreover, the resulting compactness exceeds that of a BH).
Somewhere, one has to consider 
the limited range of $\beta$ for a given mass as well, ideally by 
incorporating the constraint into the multivariate prior probability 
density used for $\Lambda,\beta,M$.

\subsection{Branch Splitting for Parametrized EOS}
\label{sec:splitting}

Our results for parametrized EOS indicate that the possibility 
of multiple stable branches needs to be taken seriously in parameter 
estimation studies using parametrized EOS. To emphasize this point,
we visualized the branch structure along a one-parametric family of
EOS connecting one of the nuclear physics EOS to the modified version
with the largest deviation from universality, as shown in 
\Fref{fig:parametrized_eos}. The upper panel shows how,
moving along this family, the single stable branch present for the 
original EOS first develops a plateau and then a second local 
maximum, thus splitting into two stable branches connected by an unstable 
one. To determine the stable and unstable branches,
we use the mass-radius criteria from \cite{Bardeen:1966:505B} 
valid for cold NS EOS.
As shown in the plot, the $M$-$R$ curve bends 
counter-clockwise at the first local mass maximum, which means that
one stable radial mode becomes unstable. At the subsequent local
mass minimum, the curve bends clockwise, meaning that the single unstable
mode becomes stable again. Finally, the second stable branch terminates 
at the second mass maximum, where the bend is counter-clockwise and
a mode becomes unstable again.

We point out that the splitting of branches leads to a conceptual 
complication in parameter estimation studies using parametrized EOS.
As shown in the middle panel of \Fref{fig:parametrized_eos}, the 
tidal deformability as function of mass becomes multi-valued where 
the mass ranges of the two branches overlap.
To set up the multivariate prior for mass and deformability, even when 
just assuming a fixed EOS together with some mass prior, one now has to chose 
a branch for certain masses. This implies a physical assumption of how 
NS form. For example, one could assume that for a given mass, a NS is always 
located on the branch with lower central density. Or one could
introduce a discrete ``branch selection'' prior instead. 
When assuming that both branches are allowed in some mass range, 
a further complication arises from the question which combinations
are realized in a neutron star binary. The corresponding prior
could just assume that each NS is randomly assigned a branch based 
on the single-NS branch prior, or it could introduce correlations 
incorporating models of binary formation channels.
When using priors for mass and EOS parameters (which are not 
independent because of the maximum mass for each EOS) the prior
for the tidal deformability is only fully determined after 
incorporating another prior for the selection of branches.
The resulting full prior for mass, EOS, and deformability might
become quite complex.

\Fref{fig:parametrized_eos} also shows that the tidal deformability for 
a given mass can differ quite strongly between two branches. When 
assuming that a NS is always 
on the the lowest-density branch available for a given mass,
it implies a discontinuity in $\Lambda(M)$. For an unequal mass 
binary, one cannot model the effective deformability anymore
by approximating $\Lambda(M)$ by a low-order Taylor-expansion.
For the EOS in \Tref{tab:eos_ns_props2} however, there 
is just one stable branch, and the slope parameter $S$ remains 
useful.

Finally, we discuss how well GW observations from BNS coalescences can
distinguish between the two stable branches. To estimate the order of magnitude
of waveform differences, we use the post-Newtonian ``TaylorF2'' model
\cite{Brown:2007jx, Damour:2000zb, Damour:2002kr} as implemented in
LALSimulation \cite{lalsuite}. The agreement between signals is quantified by a
noise-weighted inner product, or overlap \cite{Finn:1992wt}. We assume Advanced
LIGO's design sensitivity \cite{KAGRA:2013rdx, adLIGONoise} with lower cutoff
frequency at $20\,{\rm Hz}$.

\begin{figure}
\caption{Mismatch between gravitational waveforms of equal-mass BNS mergers
computed for the case that both NS are located on the high-density stable 
branch of TOV solutions and for the case that both are on the low-density 
stable branch, for an EOS that leads to two stable branches.
Further, we show the mismatch between both NS being on the same 
branch and each NS being on a different branch. 
The EOS used here is the one from \Fref{fig:parametrized_eos} with the 
largest deviation from the original one. The mismatch is shown in the 
mass range covered by both branches. In addition, we show the mismatches
between each branch and the single stable branch of the original, 
unmodified EOS.
The mismatch has been computed using the noise PSD for Advanced
LIGO's design sensitivity \cite{KAGRA:2013rdx, adLIGONoise}.
It is plotted as function of the mass of the constituent NSs.}
\label{fig:branch_mismatch}
\includegraphics[width=\columnwidth]{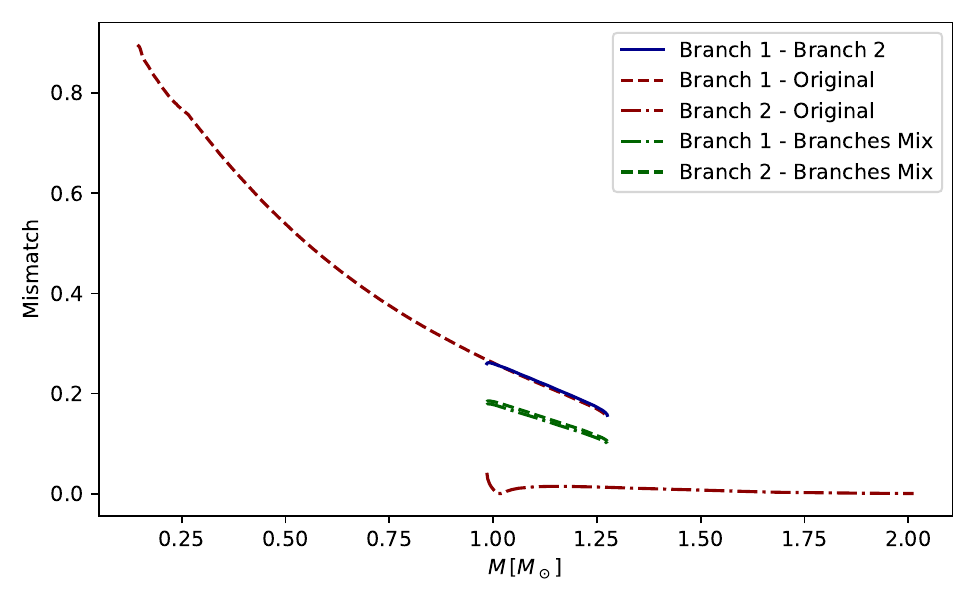}
\end{figure}

\Fref{fig:branch_mismatch} shows the mismatch (i.e., $1 - $ the overlap)
between waveforms for an equal mass BNS coalescence that only differ in 
the tidal deformability parameter. 
For each BNS waveform, we assume that both NS follow 
the same EOS and are also on the same branch.
We compare the two stable branches of the 
modified EOS shown in \Fref{fig:parametrized_eos} that deviates most from
the original one.
In addition, we compare each of those branches to the single stable 
branch of the 
original MPA1 EOS. 

Branch~1 (lower mass branch) of the modified EOS is characterized by very large tidal 
deformabilities, $\Lambda > 10^4$. These large values lead to a 
significant dephasing between signals assuming the original EOS 
and Branch~1, respectively, and also between the two branches of the 
modified EOS. This corresponds to
mismatches $> 0.2$ that are significant for any SNR above the detection
threshold. For example, $0.5 \,{\rm SNR}^{-2}$ is commonly used in the
literature as a conservative mismatch threshold for indistinguishable signals
\cite{Lindblom:2008cm, Ohme:2011zm}. Conversely, a mismatch $\sim 0.2$ would
start to become distinguishable already for SNRs $> 1.6$. Therefore,
if the EOS were known and similar to our example,
ground-based detectors would be able to
distinguish the two branches of the modified EOS example,
and also between branch~1 of the modified EOS and the original EOS.

For masses $M > M_\odot$ in \Fref{fig:branch_mismatch}, 
Branch~2 (higher mass branch) of the modified EOS yields lower
tidal deformabilities than the original EOS. The GW signals in this
case are more similar and mismatches significantly lower (between $10^{-2}$
and $10^{-4}$). The signal differences in this regime only become
distinguishable for SNRs between $\mathcal O(10)$ and $\mathcal O(100)$.

We also considered the possibility of a BNS where the two NS  
are on different stable branch of the same EOS. 
\Fref{fig:branch_mismatch} also shows the mismatch
between the case where both NS are on the same branch and the case where
each NS is on a different branch (still for equal masses).
Not surprising, the mismatches are somewhere in-between the mismatch 
between branches when both NS are on the same branch.

These are order-of-magnitude estimates for an extreme example that do not take
other parameter variations into account. Nevertheless, our considerations show
that signal differences between the different branches may become
observationally relevant in parameter estimation studies based on 
parametrized EOS.
Any such study that aims to constrain the EOS therefore needs to 
incorporate the branch structure.

\section{Summary and Outlook}
\label{sec:summary}
In this article, we collect everything required to compute 
the properties of nonrotating NS in GR. We provide novel 
analytic formulations of the tidal deformability differential 
equations that are robust when employing EOS with phase 
transitions. We also point out pifalls that occur in 
numerical implementations and how to avoid those.

The equations have been implemented in a publicly available library
\texttt{RePrimAnd}. We demonstrate the accuracy of the solution
and the robustness of the code using a wide range of example models.
In addition, we used the convergence tests to set up a model of the 
error budget that was incorporated into the library, allowing users
to specify the desired accuracy directly.

Our library makes it easy to compute NS models within C++ or Python 
code, or in an interactive Python environment such as Jupyter notebooks. 
Besides the NS functionality, the library also provides a consistent
and generic interface to the use and exchange of various EOS models 
in a transparent manner. The aim of this library is to be useful for 
applications in GW data analysis involving NS, the testing of novel 
phenomenological relations between NS properties, for mapping novel EOS 
models to NS properties, and for initial data generation in numerical 
relativity.

As a first application, we compute properties of a typical NS with
fiducial gravitational mass of $1.4 \, M_\odot$ for various EOS, 
including baryonic mass, radius, moment of inertia, and tidal deformability.
We also provide the first derivative of the tidal deformability 
with respect to mass. This might be useful in BNS GW data analysis
studies that 
reduce the EOS uncertainty to the values of tidal deformability
and its derivate at some fiducial mass.
Further, we compute the maximum mass models and their properties 
for each EOS. We also provide the corresponding EOS files for 
reference, as well as files with the TOV sequences. Those EOS 
and sequences are easily accessible through the Python interface 
of our library.

As a second application, we explore the reliability of 
universal relations between compactness and tidal deformability.
We construct EOS for which those relations are violated much
more than for nuclear physics EOS models. Those examples
are based on nuclear physics EOS but modified above $1/4$ of the 
nuclear saturation density. They show that the universal 
relations can only be employed in situations where the EOS
is already constrained to higher densities, but not for
studies aiming to constrain the EOS without prior constraints.

As a third 
application, we demonstrate how the stable branch of a NS sequence
splits into two branches when transitioning the EOS within a 
parametrized EOS family. We point out that this possibility needs
to be taken into account on the technical level, but also requires
addressing the physical model which branch NS with a given mass
would occupy in nature. Further, we show that in such situations, it
becomes infeasible to approximate the mass dependency of the tidal 
deformability using a Taylor expansion.

We show that branch splitting can cause significant differences 
in GW signals of BNS coalescence and demonstrate that those differences
are relevant for current and future ground-based GW detectors.
We note that our examples assume EOS constrains only at low densities. 
For studies that already assume EOS constraints at higher densities,
branch splitting might not be an issue.

There are some NS properties not implemented in our library, which we leave 
for future versions. Most notable examples are the oscillation 
frequencies and the spin-induced quadrupole moment. Further, it may
be useful to support more EOS types directly, such as spectral 
representations (although anything can be represented by tabulated EOS).
In general, we hope that our library will become useful for the neutron 
star community as a common infrastructure.

\acknowledgments
This work was supported by the Max Planck Society's Independent Research Group 
Program. Some computations have been carried out on the \texttt{holodeck} 
cluster at the Max Planck Institute for Gravitational Physics, Hanover.
 
\bibliographystyle{article}
\bibliography{article}

\end{document}